\documentclass[aps,prd,nofootinbib,twocolumn,superscriptaddress,preprintnumbers,balancelastpage,longbibliography]{revtex4-2}

\usepackage{multirow}
\usepackage{float}
\usepackage{graphicx} 
\usepackage{url}
\usepackage{graphicx}
\usepackage{amsmath,amssymb,amstext,amssymb,amsfonts,amsthm}
\usepackage{aas_macros}
\usepackage{hyperref}
\usepackage{appendix}
\usepackage{physics}
\usepackage{soul}
\usepackage{comment}
\usepackage{color}
\usepackage[normalem]{ulem}
\usepackage{lipsum}
\usepackage{capt-of}

\newcommand{\bea}{\begin{eqnarray}}
\newcommand{\eea}{\end{eqnarray}}
\newcommand{\vth}{v_{\rm th}}
\newcommand{\df}{\delta \!  f}
\newcommand{\nn}{\nonumber}
\newcommand{\vthe}{v_{\rm th}^{e}}
\newcommand{\vthi}{v_{\rm th}^{i}}

\usepackage{todonotes}

\definecolor{deepgreen}{rgb}{0.2,0.8,0.2}

\definecolor{deepblue}{rgb}{0.2,0.2,0.8}

\definecolor{deepred}{rgb}{0.8,0.2,0.2}

\begin{document}
\title{No cosmological constraints on dark photon dark matter from resonant conversion: Impact of nonlinear plasma dynamics
}

\author{Anson Hook}
\email{hook@umd.edu}
\affiliation{Maryland Center for Fundamental Physics, University of Maryland, College Park, MD 20742, USA}

\author{Junwu Huang}
\email{jhuang@perimeterinstitute.ca}
\affiliation{Perimeter Institute for Theoretical Physics, 31 Caroline St.~N., Waterloo, Ontario N2L 2Y5, Canada}

\author{Mohamad Shalaby}
\email{mshalaby@perimeterinstitute.ca}
\altaffiliation{Horizon AstroPhysics Initiative (HAPI) Fellow}
\affiliation{Perimeter Institute for Theoretical Physics, 31 Caroline St.~N., Waterloo, Ontario N2L 2Y5, Canada}
\affiliation{Waterloo Centre for Astrophysics, University of Waterloo, Waterloo, ON N2L 3G1, Canada}
\affiliation{Department of Physics and Astronomy, University of Waterloo, Waterloo, ON, N2L 3G1, Canada}

\date{\today}

\begin{abstract}
We revisit and invalidate all dark photon dark matter constraints from resonant conversion of dark photons into photons (plasmons) in the early universe. These constraints rely on the resonant transfer of a substantial portion of the dark photon energy density into the SM plasma, heating the plasma in the process. We demonstrate that this resonant transfer saturates because of plasma nonlinearities. Dark photon dark matter resonantly converts into $k \simeq 0$ Langmuir waves in the early universe electron-ion plasma. Once the Langmuir-wave energy approaches the thermal energy of the plasma, nonlinear effects driven by the ponderomotive force become significant. In particular, we show using dedicated Particle-in-Cell simulations that large-amplitude $k = 0$ Langmuir waves excite higher-k Langmuir and ion acoustic waves, producing strong spatial variations in density and plasma frequency. These inhomogeneities suppress further resonant conversion, limiting the deposited energy to about the thermal energy of the electrons at the time of conversion, orders of magnitude below observable cosmological thresholds. Consequently, the dark photon dark matter constraints are weaker by factors of $3000$ to $10^7$ across ten orders of magnitude in dark photon mass.
\end{abstract}
\maketitle

\paragraph*{\bf Introduction.---} 

The existence of dark matter (DM) in our universe offers the strongest evidence for the incompleteness of the Standard Model (SM)~\cite{1970ApJ...159..379R,Bergstrom:2000pn,Bertone:2004pz}. Despite overwhelming gravitational evidence for DM~\cite{1970ApJ...159..379R,Planck:2018vyg}, we have yet to measure nongravitational interactions between DM and SM. 

A particularly well-motivated class of candidates for DM is light bosonic DM, which emerges at low-energy in extradimensional theories, such as string theory~\cite{Svrcek:2006yi,Arvanitaki:2009fg,Goodsell:2009xc,Kaluza:1921tu,Klein:1926tv} and can be produced as DM gravitationally~\cite{Preskill:1982cy,Redondo:2008ec,Nelson:2011sf,Graham:2015rva,Gorghetto:2020qws}.
Most searches for light bosons, in astrophysics, cosmology, and laboratory, are based on (resonant) conversion to SM photons. In this letter, we focus on one example of these conversions between dark photon dark matter (DPDM) and the SM photon in cosmology, though similar considerations can apply to a variety of light bosons and astrophysical environment. 

The dark photon couples to the SM through kinetic mixing between the dark photon field strength $F'_{\mu\nu}$ with SM photon field strength $F_{\mu\nu}$~\cite{Holdom:1985ag,Okun:1982xi}:  
\bea
\delta \mathcal{L} \supset \frac{\epsilon}{2} F^{\mu\nu} F'_{\mu\nu} \qquad \Rightarrow \qquad \epsilon e A' J_{\rm E \& M} \, .
\eea
This kinetic mixing term gives the dark photon a coupling to the SM electromagnetic current $J_{\rm E\& M}$, suppressed by an unknown kinetic mixing parameter $\varepsilon$. DPDM with this coupling is the target for many proposed and currently running experiments~\cite{Chaudhuri:2014dla,Baryakhtar:2018doz,SuperCDMS:2019jxx,FUNKExperiment:2020ofv,SENSEI:2020dpa,Chiles:2021gxk,Cervantes:2022yzp}.  In addition to terrestrial experiments, there are also constraints from astrophysics and cosmology~\cite{Fixsen:1996nj,Mirizzi:2009iz,Kunze:2015noa,McDermott:2019lch,Caputo:2020rnx,Caputo:2020bdy,Pirvu:2023lch,Siemonsen:2022ivj,McCarthy:2024ozh,Chluba:2024wui}.

The most stringent limit on DPDM over ten orders of magnitude in mass from about $10^{-14}$ eV to $10^{-4}$ eV comes predominantly from the conversion of DPDM into a plasma excitation in the early universe electron ion plasma, heating the plasma in the process~\footnote{We adopt the natural units of $\hbar = c = k_B = 1$.}~\cite{Arias:2012az,Caputo:2021eaa}. Within perturbation theory, the probability of energy transfer is given by the Landau-Zener formula~\cite{Arias:2012az}
\begin{equation}\label{eq:LZ}
    P_{A'\rightarrow \gamma} = \pi \varepsilon^2 m_{A'}\left.\left(\frac{{\rm d} \log[\omega_p^2]}{d \chi}\right)^{-1} \right|_{\omega_p = m_{A'}} \sim \varepsilon^2 \left(\frac{m_{A'}}{H}\right),
\end{equation}
where $\chi$ is the line of sight distance, $\omega_p = \sqrt{e^2 n_e/m_e}$ is the plasma frequency and $m_{A'}$ the dark photon mass~\footnote{The production mechanisms of DPDM only apply if $m_{A'}$ is from a Stuckelberg mechanism~\cite{Stueckelberg:1938hvi,Ruegg:2003ps} due to string formation~\cite{East:2022rsi,Cyncynates:2023zwj,Cyncynates:2024yxm}.}, while $H$ denotes the Hubble scale at the time of the resonant transfer~\footnote{Except for during reionization and recombination where the time dependence of the ionization fraction $X_e$ is more important.}. 
Such an effect has been used to put limits on the DPDM parameter space with $N_{\rm eff}$ observations at high mass~\cite{Arias:2012az}, spectral distortions in the intermediate mass~\cite{Arias:2012az}, and dark ages reionization, Helium reionization and similar considerations at low mass~\cite{Caputo:2020rnx,Caputo:2020bdy,McDermott:2019lch,Witte:2020rvb}. 

These limits are obtained by assuming that a substantial portion of the total DPDM energy density is transferred into the SM plasma. For example, the limit from $\mu$ and $y$-distortion corresponds to when about $\mathcal{O} (10^{-4})$ of the total energy density in radiation is injected into the SM plasma~\cite{Chluba:2016bvg}, however, as we will show, at most about $\mathcal{O} (10^{-8})$ of the total radiation energy density can be injected into the SM plasma before the approximation employed in~\cite{Arias:2012az,Caputo:2020rnx,Caputo:2020bdy,McDermott:2019lch,Witte:2020rvb} breaks down. The electron ion plasma goes nonlinear, and the resonant conversion shuts off. In this letter, we will first clarify when the commonly used linear approximation breaks down and identify the leading nonlinear effects analytically.  Afterwards, we will use a Particle-in-Cell (PIC) simulation to demonstrate how these nonlinearities shut off the resonant conversion and invalidate many DPDM constraints.

\paragraph*{\bf Heuristic arguments and analytical results.---}
Kinetically mixed DPDM behaves like an electric field that oscillates with frequency $\omega_{A'} = m_{A'}$ and amplitude $\varepsilon E' \sim \varepsilon \sqrt{\rho_{\rm DM}}$, where $\rho_{\rm DM}$ is DM energy density.  Because DM is extremely non-relativistic, the $k/\omega$ of this electric field is extremely small.  For example, many DPDM production mechanisms produce dark photons with $\gamma \sim 1$ when $H \approx m_{A'}$~\cite{Graham:2015rva,Amin:2022pzv}.  In this case, at the time of conversion, $k/\omega \sim \sqrt{ H/\omega_p} \ll 1$. Therefore, we describe the response of a plasma to a $k \approx 0$, $\omega = m_{A'}$ oscillating external electric field acting on resonance $\omega_p = m_{A'}$, though the following discussions apply more generally.

The breakdown of the perturbative expansion can be identified by inspecting the Vlasov equations in 1+1D. DPDM, through kinetic mixing, drives Langmuir waves (longitudinal electrostatic oscillations) in the SM plasma.  Qualitatively, these longitudinal plasma oscillations drive local bulk motion in electrons (relative to ions) leading to a charge current $ j \approx \omega_p E $, with $E$ the electric field strength in this plasma. Perturbation theory breaks down when this current is larger than $ e n_e v^e_{\rm th}$, that is, when the collective motion of the electrons exceeds the random thermal motion of the electrons, where the electron thermal speed is defined as $v^e_{\rm th} \equiv \sqrt{T_e/ m_e}$.
This occurs when (see App.~\ref{app:plasmababy} for more details)
\begin{equation}
    e E \approx \frac{e^2 n_e v^e_{\rm th}}{\omega_p} = m_e \omega_p v^e_{\rm th},
\end{equation}
which is also when the electric field energy equals the initial thermal energy of the electron. The ratio of $e E/m_e \omega_p$ is defined as the quiver velocity $v_q$, and perturbation theory breaks down when $v_q/v^e_{\rm th}$ approaches unity~\cite{1985PhR...129..285Z,Schekochihin2017Houches}.

A leading nonlinear effect in this system comes from the ponderomotive force $\vec{F}_p = -\nabla \Phi_p$, where the ponderomotive potential is given by~\cite{1987PhR...145..319K} 
\begin{equation}
    \Phi_p(x) = \frac{e^2}{4 m_e \omega^2} \hat{E}^2 (x),
\end{equation}
with $\hat{E}$ the amplitude of the electric field. This force is proportional to $(e \hat{E})^2$ and acts on the ions and electrons in the same direction. This pushes both species towards regions with low wave amplitude, leading to the spatial variation of the total density $n_{\rm tot} = n_e + n_p$. If the electric field were not to dissipate, this would equilibrate to $n_{\rm tot} \propto \exp[-\Phi_p/T] = \exp[-v_q^2/4 {v^e_{\rm th}     }^2]$.
This means that the ratio $v_q/v^e_{\rm th}$ is not just the perturbative expansion parameter but also a measure of nonlinearity.  A spatial variation of $n_{\rm tot}$ change the resonant frequency and terminate resonant conversion.

For $v_q/v^e_{\rm th}\lesssim1$, a comprehensive review of how nonlinear effects take action dynamically can be found in~\cite{1985PhR...129..285Z,1997RvMP...69..507R}.
Whereas DPDM resonantly converts to Langmuir waves with $k\approx0$, higher $k$ modes get excited through a variety of instabilities, including the modulational instability
and the electrostatic decay instability~\cite{1985PhR...129..285Z,1997RvMP...69..507R}.
The growth rates of various instabilities, scaling as powers of $v_q/v^e_{\rm th}$, can be found with the Zakharov equations (where the effect of the ponderomotive force is apparent~\cite{1985PhR...129..285Z,1997RvMP...69..507R}) and has been confirmed numerically and experimentally~\cite{1999PhRvL..83.2965K,1991PhRvL..66.1970D,PhysRevLett.56.838,1988PhRvL..60..581R}. Of particular interest is the case of a $k=0$ mode exciting $k \ne 0$ modes, where the instability, the super- and sub-sonic modulational instability, resembles parametric resonance~\cite{ivanov1967quasilinear,1972JETP...35..908Z}. 

More importantly, the excitation of ion acoustic waves leads to spatial variations in ion density, and consequently in the total density $n_{\rm tot}$ and the plasma frequency, on a comparably longer timescale~\cite{Schekochihin2017Houches}. The ion acoustic wave has a dispersion of $\omega \approx k \, c_s$ for $k \lambda_{De} \ll 1$, where $\lambda_{De} = v_{\rm th}^e/\omega_p$ is the electron Debye length and $c_s \equiv \sqrt{(T_e+T_i)/m_i}$ is the sound speed.
The ion acoustic wave can be excited by the electrostatic decay instability~\cite{1972JETP...35..908Z,1985PhR...129..285Z} as well as the ion acoustic instability~\cite{1961PhFl....4..139F,Schekochihin2017Houches} depending on $v_q/v^e_{\rm th}$. As we will show with simulations, resonant conversion between DPDM and photon is inhibited by the formation of ion acoustic waves and cannot proceed before they damp, which is extremely slow~\cite{Schekochihin2017Houches} when a steady state is reached with $T_e \gg T_i$.  

\paragraph*{\bf Numerical results.---} Although the analytical results summarized in the previous section provide ample evidence for the onset of instabilities during the resonant conversion from the DPDM to the low $k$ Langmuir wave, numerical studies are needed to understand the nonlinear long-term behavior of the system, particularly in the cases where $v_q > \vthe$.
Therefore, we run Particle-in-Cell simulations to study these effects.
To do so, we adapt the SHARP code \cite{sharp,sharp2} to simulate the energy transfer between the DPDM and the SM plasma with 1+1D PIC simulations (see App. \ref{app:sim}). The 1+1D setup is able to capture the linear evolution and the most important nonlinear effects because the plasma is driven at $k\simeq 0$ by the DPDM, and the higher $k$ modes are produced due to local drift of electrons relative to ions through instabilities that are largest for $k$ that are aligned with this relative drift~\cite{bret2010}. Consequently, the fastest-growing modes will be aligned with the relative drift direction, and the system is effectively 1+1D.

\begin{figure}
    \centering
    \includegraphics[width=1\linewidth]{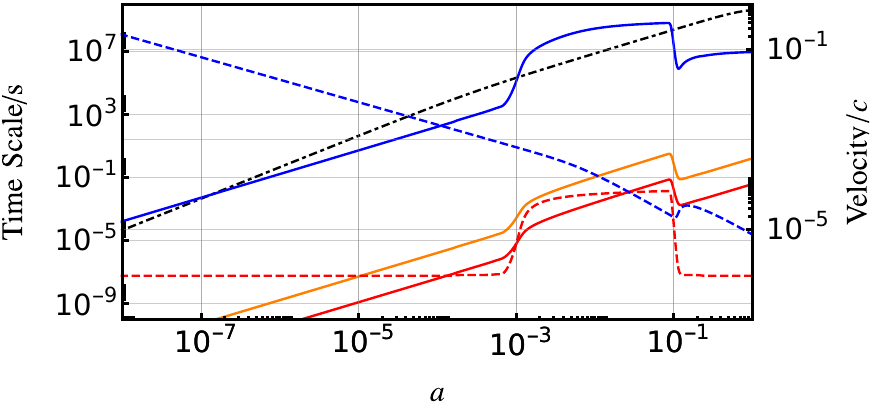}
    \caption{The important time scales (Left axis) and velocities (Right axis) in our study as a function of scale factor $a$. The red and orange solid lines show inverse electron and ion plasma frequency, the blue solid line show the electron ion energy exchange time, while the black dot-dashed line shows the duration of the resonance $\varepsilon/H$ for $\varepsilon = 10^{-8}$. The blue dashed line shows the electron thermal speed, while the red dashed shows the DPDM quiver velocity $v_q^D$ also for $\varepsilon=10^{-8}$. }
    \label{fig:velocity}
\end{figure}

As shown in Fig.~\ref{fig:velocity}, the duration of the Landau-Zener transition $\varepsilon/H$ is typically $10^{10}$ periods of plasma oscillation, which is well beyond the duration of any simulation. Therefore, to demonstrate the effect of nonlinearity, we perform two simulations: the resonant transfer between the DPDM with $\omega = \overline{\omega_p}$ and $k=0$, where $\overline{\omega_p}$ is the plasma frequency averaged over the whole simulation box, when both quantities are time independent; a Landau Zener transition with $\overline{\omega_p}/m_{A'}$ changing much faster than in realistic scenarios. 

Similarly, in a realistic situation, the DPDM driving field has amplitude 
\begin{equation}
    v_q^{\rm D} \equiv \frac{\varepsilon e E'}{\omega_p m_e} =  \varepsilon\left(\frac{2\rho_{\rm DM}}{X_e \rho_e}\right)^{1/2} \simeq 10^{-6} \left(\frac{1}{X_e }\right)^{1/2} \left(\frac{\varepsilon}{10^{-8}}\right),
\end{equation}
where ${X_e }$ the electron ionization fraction, whereas the electron thermal speed is shown in Fig.~\ref{fig:velocity}~\cite{2011JCAP...07..034B}. To best approximate the parameters of interest, we simulate a variety of $v_q^{\rm D}/v_{\rm th}^e$ ratios ranging from $0.1$ to $10^{-3}$, and show the two qualitatively different regimes of evolution~\footnote{Videos of these simulations can be found at \cite{Video_DPDM}}.

\begin{figure}[t!]
    \centering    \includegraphics[width=\linewidth]{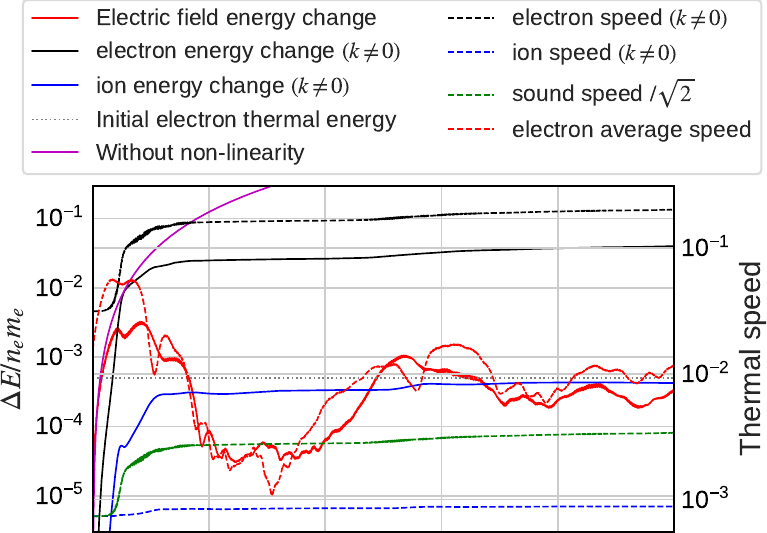}
    \\
    \vspace{0.05cm}
    \includegraphics[width=\linewidth]{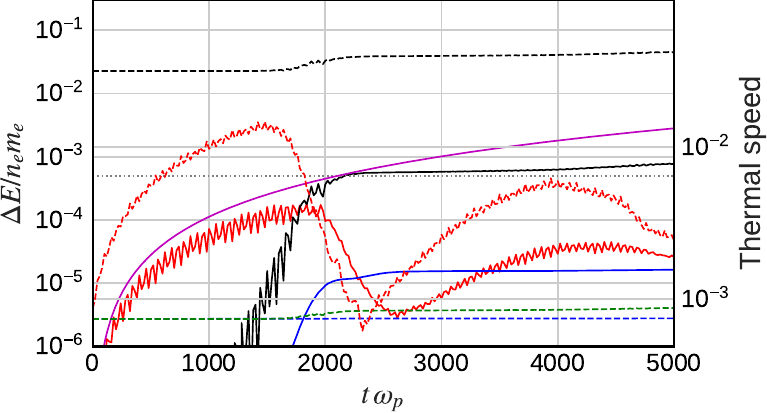}
    \caption{Numerical results for resonant conversion with $\overline{\omega_p} = m_{A'}$. The solid lines (left axis) show the various energy densities ($\Delta E$)
    as a function of time,
    while the dashed lines (right axies) show the various thermal speeds
    as a function of time. The dotted gray line shows the initial thermal energy in the system as a comparison. The purple solid line shows the naive expectation for the growth of electron energy in linear theory of $t^2$, while the red solid line shows the actual evolution in PIC simulations. The upper panel is for a strong DPDM field with $v_q^D/v_{\rm th}^e = 0.03$ while the lower panel is for a weak DPDM field with $v_q^D/v_{\rm th}^e = 10^{-3}$.
    }
    \label{fig:Efield}
\end{figure}

\paragraph{Resonant conversion} Let us first discuss the results for $ m_{A'}= \overline{\omega_p}$. As shown in Fig.~\ref{fig:Efield},~\ref{fig:panels} and~\cite{Video_DPDM}, energy is transferred from DPDM to the $k=0$ Langmuir wave at the beginning of evolution. During this period, the electric field grows linearly with time, and the energy density $\omega_p^2t^2$ (see App.~\ref{app:plasmababy}), while perturbations at higher $k$ grow due to the ponderomotive force. After this initial period, two qualitatively different behaviors exist.   

Let us start with the {\it slow growth regime} ($v_q^{\rm D}/v_{\rm th}^e \lesssim (m_e/m_p)^{1/2}/2 \approx 0.01$), shown in the lower panel of Fig.~\ref{fig:Efield} and~\cite{Video_DPDM}. The solid red line ($\langle E^2(t) \rangle/2$) grows while oscillating at the plasma frequency, closely following the linear prediction (purple line) until nonlinear effects and $k \ne 0$ modes become important. As we explained earlier, nonlinear effects become important when $v_q/\vthe \rightarrow 1$. In Fig.~\ref{fig:Efield}, this manifests itself as the average electron speed, roughly the quiver velocity $v_q$ (red dashed line) approaching $\vthe$ (black dashed line). At this point, the red line deviates from the purple, and the black line (the energy stored in the $k\neq 0$ modes) increases.
At late time, the black line matches where the red line was, indicating that $\mathcal{O}(1)$ of the energy has been transferred into the higher $k$ modes and thermal energy of the electrons. The clear departure from the purple line demonstrates that energy deposition has gone off-resonance and effectively stops.

The physics behind these numerical features is as follows.  For $v_q^{\rm D}/v_{\rm th}^e \lesssim 0.01$, the timescale for the growth of the $k=0$ Langmuir wave to $v_q/v_{\rm th}^e\simeq 1$ is much longer than that for the growth and oscillations of the ion acoustic wave. In this case, the nonlinearities lead to a slow down of the resonant transfer before the energy density stored in the Langmuir wave becomes comparable to the initial thermal energy of the electrons, roughly $(v_q/v_{\rm th}^e)^2$. 
This initial slow down can be described semi-analytically by solving the Zakharov equations~\cite{1985PhR...129..285Z}.
The high-$k$ mode of the Langmuir wave is excited by the sub- and super-sonic modulational instability, while the ion acoustic wave is excited by
the electrostatic decay instability, which also populates a variety of $k$ ranging from $\omega_p^i$ to $\lambda_D^{-1}$ at the same time~\cite{Video_DPDM}. 
Note that these density variations are on very small scales, and shall not be confused with the large scale density variations in cosmology discussed in~\cite{Caputo:2020rnx}.
The amplitude of the electric field oscillates and saturates before $v_q/v_{\rm th}^e\simeq 1$ and the total energy transfer is about the electron thermal energy.

We now turn to the {\it fast growth} regime ($v_q^{\rm D}/v_{\rm th}^e \gtrsim 0.01$) shown in the top panel of Fig.~\ref{fig:Efield} and~\cite{Video_DPDM}.  In this regime, the energy injected grows rapidly and the electron dynamics becomes non-perturbative almost instantaneously.  After a time of $\sim 1/\omega_p^i$, the ions respond, nonlinearities become important, and the growth halts.

The physics behind the fast growth regime is that the perturbations of the electron oscillations grow first at large $k$ through the supersonic modulational instability~\cite{1985PhR...129..285Z,1997RvMP...69..507R}, whereas ion acoustic oscillations grow on a much longer time scale through the ion acoustic instability. 
In this case, the electric field energy can grow past the thermal energy of the system, and saturate as the ion acoustic perturbations grow. The Langmuir waves heat up the electrons in this process, and the electron temperatures grow to $\mathcal{O}(10^2)$ of the initial temperature of the system. The system saturates and reaches a steady state.  

\begin{figure}
    \centering
   \includegraphics[width=1\linewidth]{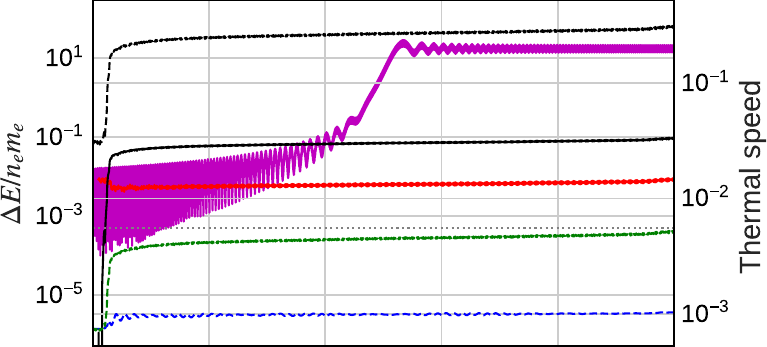}    
    \\
    \vspace{0.05cm}
 \includegraphics[width=1\linewidth]{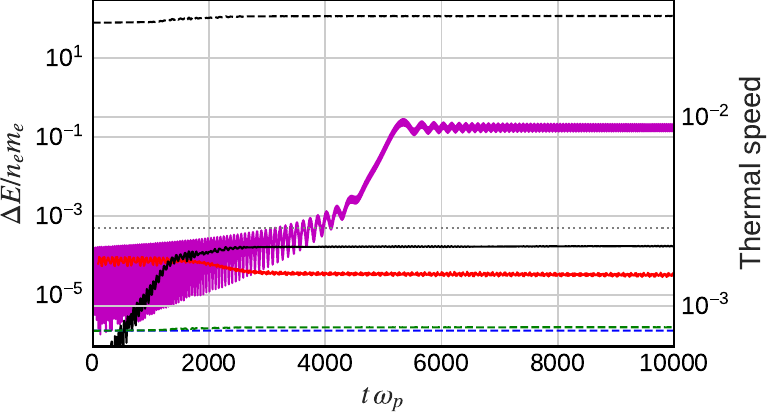}
    \caption{\label{fig:LZcase}%
    Numerical results for a Landau-Zener transition with $v_q^D/v_{\rm th}^e = 1$ (top) and $v_q^D/v_{\rm th}^e = 0.1$ (bottom). The purple solid line shows the naive expectation for the growth of electron energy in linear theory (see Eq.~\ref{eq:LZ}), while the red solid line shows the actual evolution in a PIC simulation. Color coding is the same as in Fig.~\ref{fig:Efield}.
    }
\end{figure}

\paragraph{Landau Zener Transition} 
While illustrative, the preceding resonant case is significantly more efficient than a Landau–Zener transition. Another key distinction is that, even away from the exact resonance, both Langmuir and ion acoustic waves can be excited, leading to two main consequences. First, nonlinear effects trigger early off resonance energy transfer (see the purple and black lines in Fig.~\ref{fig:LZcase}). Second, the off-resonant excitation makes it more difficult for the system to overshoot, as observed in the {\it fast growth} regime of the previous cases.

Here we present simulations that correspond to $\omega_p/H \approx 7 \times 10^4$. As it is difficult to numerically simulate a plasma with a changing plasma frequency, we instead facilitate this level crossing with a $m_{A'} (t)$ that grows with time.  This approach works since, as long as the variation with time is small, the LZ resonant transfer primarily depends on the ratio of $\omega_p/m_{A'}$ and its time derivative. Additionally, due to computational run time, the simulated parameters are larger than the extreme ratio of $\omega_p/H$ present for DM (see Fig.~\ref{fig:velocity}). 

In Fig.~\ref{fig:LZcase}, we show a comparison between the resonant transfer with the linear theory treatment and our numerical results, where it is clear that nonlinearity kills the resonant transfer period around $\omega_p t = 5000$ almost entirely. Furthermore, because the initial $v_q/\vthe$ is chosen so large, if this was the resonant case, it would cause significant overshooting.  Instead, the kinetic energy of the electrons is not even changed by a factor of 2.

\paragraph{Ion acoustic wave damping}
Before we close this section, let us return to the long term stability of the ion acoustic wave. Ion acoustic waves damp quickly on ions in a thermal electron ion plasma~\cite{Schekochihin2017Houches}. 
However, this damping rate on ions would be exponentially suppressed if the sound speed is much larger than the ion thermal speed, by a factor of $\exp[ - ({c_s}/{v_{\rm th}^i})^2/2] \approx \exp[ - ({T_e}/2T_i)]$ when $T_e \gg T_i$, while the damping rate on electrons would be polynomial suppressed if the electrons were thermalized with themselves. The final steady state we find contains a mixture of Langmuir waves and ion acoustic waves with electrons about $30$ times hotter than the ions (see more details in App.~\ref{app:simresult}). As shown in Fig.~\ref{fig:Efield} and even more so in~\cite{Video_DPDM}, this final state is stable enough to prevent any transfer of energy until $\omega_p t = 8\times 10^4$, the duration of the longest of our simulations, which is already approaching when collisions between particles shall become relevant (see more details in App.~\ref{app:simresult}). While we can only simulate for long enough to see this eventual saturation for relatively large $v_q^D/v_{\rm th}^e$ due to limited simulation time, we expect this to hold for all cases of interests. 

On cosmological time scales, the electrons and ions equilibrate at least due to collisions. This time scales is set by the collisions between the electrons and the environment, most importantly ions, with (momentum exchange) rate~\cite{1950PhRv...80..230C}
\begin{equation}\label{eq:nuei}
    \nu_{ei} = \frac{4 \sqrt{2 \pi}n_e\alpha^2 \log \Lambda}{3m_e^2} \left( \frac{m_e}{T_e}\right)^{3/2},
\end{equation}
where $\log \Lambda = \log (4\pi T_e^3/\alpha^3 n_e)^{1/2} \approx 20$ in the early universe, with $\alpha$ the fine structure constant. The electron–ion energy exchange time~\cite{1950PhRv...80..230C,1962pfig.book.....S,2016PhPl...23f4504T} 

\begin{equation}
    \tau_{ei} = \frac{3m_e m_p}{8 \sqrt{2 \pi}n_e\alpha^2 \log \Lambda} \left( \frac{T_e}{m_e}\right)^{3/2},
\end{equation}
is the important time scale (see Fig.~\ref{fig:velocity}), which is about $10^4 \,{\rm s}$ at recombination, or about $10^{-9}$ compared to the Hubble scale at the time\footnote{This is smaller than  $1/\nu_{ei}$ by the ion-electron mass ratio.}. 
The time scale $\tau_{ei}$ sets the rate of the thermalization between the electrons and ions, and as a result, the time scale over which the ion acoustic wave damp. Dedicated simulations of a collisional plasma might be useful to understand this final state.

\paragraph*{\bf Dark Photon Dark Matter constraints.---} Our numerical results indicate that resonant conversion can inject up to $\mathcal{O}(100)$ times the electron thermal energy, with typical cases depositing $\mathcal{O}(1)$. This piddling amount of energy deposition essentially removes all resonance conversion constraints on DPDM.  e.g. spectral distortion bounds require energy deposition of $10^{-4} \rho_\gamma$, whereas 100 times the electron thermal energy is only $10^{-8} \rho_\gamma$.

\begin{figure}
    \centering
    \includegraphics[width=0.9\linewidth]{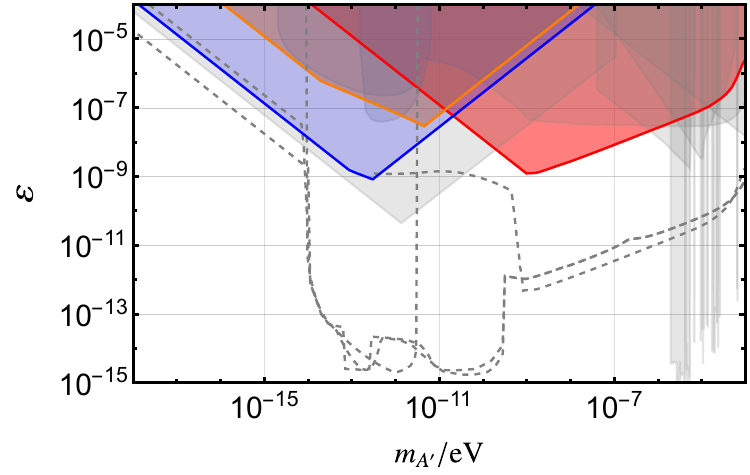}
    \caption{Updated Dark Photon Dark Matter Limits. The gray shaded regions are constraints from a variety of astrophysical and lab searches~\cite{Caputo:2021eaa}, while the color shaded regions are the cosmological constraints from early universe considerations (spectral distortion and $N_{\rm eff}$ in red) and late universe considerations from Dark Ages (orange) and Lyman-$\alpha$ forest (blue). 
    Constraints on dark photons also arise from vector superradiance with gravitational wave measurements~\cite{LIGOScientific:2025csr,Aswathi:2025nxa}. 
    The previous invalidated constraints are shown as gray dashed lines~\cite{Arias:2012az,Caputo:2020rnx,Caputo:2020bdy,McDermott:2019lch,Witte:2020rvb}.
    }
    \label{fig:money}
\end{figure}

Such an energy injection is smaller than the non-resonant heating studied in~\cite{Dubovsky:2015cca} for all $m_{A'}$ and $\varepsilon$ that is not already constrained. The non-resonant heating rate scales as $\varepsilon^2\nu_e/H$, instead of $\varepsilon^2 \omega_p/H$ as in Eq.~\ref{eq:LZ}
and therefore leads to constraints that are weaker than~\cite{Arias:2012az} by a factor of more than 3000 before recombination, as shown in Fig.~\ref{fig:money} as the red shaded region for $m_{A'} \gtrsim 10^{-9}\, {\rm eV}$. 

After recombination, the constraints on exotic energy injection come from measurements that are sensitive to changes of ionization fraction~\cite{Witte:2020rvb,Caputo:2020bdy}, as well as the temperature of the gas~\cite{Liu:2020wqz}. 
Unlike what was considered before in~\cite{Witte:2020rvb}, the energy injection from non-resonant conversion cannot be treated as instant injection, and the energy injection rate significantly decreases as  $ T_e^{-3/2}$ as the temperature of the electron increases. 
We compute the terminal temperature as a function of $\varepsilon$ and $m_{A'}$ and impose that the universe cannot heat up to temperatures at which collisional reionization occurs during dark ages ($\sim 10 \,{\rm eV}$ for $20<z<500$), or temperatures that lead to line broadening of Lyman-$\alpha$ forest ($\sim 0.8 \,{\rm eV}$ for $2<z<6$) in all allowed parameter space (see App.~\ref{app:constraints} for details). The tentative limits from after recombination are shown as the orange and blue shaded region in Fig.~\ref{fig:money}. We leave a careful study of the effects of non-resonant heating during recombination to future work. 

We caution that the above-mentioned instabilities can also invalidate constraints stemming from resonant conversion of axion DM and DPDM into electromagnetic waves around a variety of dilute (magnetized) plasma near astrophysical bodies~\cite{Hook:2018iia,An:2024wmc}, or dark stars~\cite{Fox:2025tqa}. We leave studies involving magnetized plasma to future work. 

\paragraph*{\bf Conclusion.---} In this letter, we show that the resonant conversion from DPDM to plasma excitations shuts off before a substantial amount of energy can be transferred to the plasma due to plasma instabilities. 
Our analytical and numerical results suggest that these nonlinear effects become important as the energy transferred approach the initial thermal energy (pressure) in the electron ion plasma. The excitation of high $k$ Langmuir wave, and more importantly, ion acoustic wave leads to spatial variation of the plasma frequency, and termination of resonant growth in numerical simulations.

The inability to resonantly transfer a substantial amount of energy leads to a much weaker constraints on DPDM. Over ten orders of magnitude in mass between $10^{-14}$ to $10^{-4}\,{\rm eV}$, the limit on $\varepsilon$ is weakened by at least a factor of 3000. This highlights that non linear plasma effects can play a significant role in our understanding of the evolution of light DM in the universe.

\acknowledgments
We thank Will DeRocco for providing us with many useful tools and enlightening discussions.
We thank Luciano Combi, Neal Dalal, Michael Fedderke, Hongwan Liu, Wenzer Qin and Zach Weiner for helpful discussions.

Research at Perimeter Institute is supported in part by the Government of Canada through the Department of Innovation, Science and Economic Development and by the Province of Ontario through the Ministry of Colleges and Universities.
M.S. receives additional support through the Horizon AstroPhysics Initiative (HAPI), a joint venture of the University of Waterloo and Perimeter Institute for Theoretical Physics.
This work was supported by the North-German
Supercomputing Alliance (HLRN) under project bbp00078.  AH is supported by NSF grant PHY-2514660 and the Maryland Center for Fundamental Physics.

\begin{appendix}

\section{A lightning fast introduction to Plasma Physics}\label{app:plasmababy}

In this appendix, we seek to provide a quick introduction for particle physicists to the very complex and rich field of plasma physics. We apologize to the experts and refer readers to~\cite{Schekochihin2017Houches,1985PhR...129..285Z,1997RvMP...69..507R} for a much more in-depth and precise review to learn about this incredible field.  While much of what is done in the main text requires numerical simulations, many of the features are also present in the simpler toy examples that we will consider below.  The starting point of all plasma physics calculations are the Vlasov equation and Maxwell's equations
\bea
\label{Eq: vlasov-maxwell}
\frac{\partial f_\alpha}{\partial t} &+& \vec v \cdot \frac{\partial f_\alpha}{\partial \vec x} + \frac{q_\alpha}{m_\alpha} ( \vec E + \vec v \times \vec B ) \cdot \frac{\partial f_\alpha}{\partial v} = 0 \\
\vec \nabla \cdot \vec E &=& \sum_\alpha q \int d^3v\, f_\alpha \nn \\
\vec \nabla \cdot \vec  B &=& 0\nn \\
\vec \nabla \times \vec B &=& \frac{\partial \vec E}{\partial t} + \sum_\alpha q \int d^3v\, \vec v f_\alpha \nn \\
\vec \nabla \times \vec E &=& - \frac{\partial \vec B}{\partial t} \, .\nn
\eea
Throughout this section, $\alpha$ will label particle species, typically electrons and ions.  When there are multiple sub/superscripts, they will often indicate which type of species the quantity is defined for.  For example, $\omega_p^e$ ($\omega_p^i$) will be the plasma mass for the electron (ion). 
On the other hand, when there are no explicit indication, e.g. $\omega_p$, it will be understood that the quantity is for the electron.  Additionally, for simplicity we will limit ourselves to the electrostatics limit in 1+1D, meaning that we can drop all B fields and vector indices.

\subsection{Perturbation Theory}

When doing perturbation theory in plasma physics, one typically does perturbation theory in $e E/m_e \omega_p \vth$.  To see this, we will provide a simple example reminding readers where the plasma frequency comes from and how $k\ne 0$ modes are necessary for nonlinearities to arise.

Langmuir waves are longitudinal oscillations of the electron plasma and electric field.  This can somewhat colloquially be called the longitudinal mode of the photon after it acquires a plasma mass.

As a warm-up, let us calculate the frequency of oscillation of a $k=0$ Langmuir wave, which has frequency equal to the plasma mass ($\omega =\omega_p$).  Let us take there to be a {\it known} electric field $E_z = E \cos (\omega t)$ and solve self consistently for the {\it unknown} phase space distribution, $f$, using perturbation theory.  The Vlasov equation becomes
\bea
\partial_t f =\frac{e}{m_e} E_z \partial_v f \, .
\eea
For simplicity we will consider only the electrons for now, but it is simple to include heavier protons/ions if need be.  The x and y directions are unaffected by the electric field and will be ignored, and the z direction label will be omitted for clarity of notation in the following discussions.  We will solve for the phase  space distribution perturbatively using
\bea
f = \sum f_i \qquad f_i \propto E^i \, ,
\eea
and in the process discover in what dimensionless object is perturbation theory being done with.  Solving to leading order in perturbation theory, we find
\bea
f_0 &=& n_e \sqrt{\frac{m}{2 \pi T}} e^{- \frac{m v^2}{2 T}}  \equiv \frac{n_e}{\sqrt{2 \pi } \vth} e^{- \frac{v^2}{2 \vth^2}} \nn \\
f_1 &=& \left ( \frac{e E}{m_e \omega }\right ) \left ( \partial_v f_0 \right )  \sin (\omega t ) \nn \\
&=& - \left (  \frac{e E}{m_e \omega \vth} \right )  \left (  \frac{v}{\vth} \right ) f_0  \sin (\omega t ) \nn \\
f_2 &=& \frac{1}{2} \left ( \frac{e E}{m_e \omega }\right )^2 \left ( \partial^2_v f_0 \right )  \sin^2 (\omega t ) \nn \\
&=& \frac{1}{2} \left ( \frac{e E}{m_e \omega \vth}\right )^2 \left (   \frac{ v^2}{\vth^2} - 1  \right ) f_0  \sin^2 (\omega t ) \, .\nn
\eea
While not important, we have taken the unperturbed distribution to be that of thermal equilibrium with rms velocity in the z direction of $\vth \equiv \sqrt{T/m_e}$.  From this expression, we see a general lesson in plasma physics, the small number in perturbation theory is~\cite{1985PhR...129..285Z}
\bea
\frac{e E}{m_e \omega \vth} \ll 1 \, .
\eea
Once, this number becomes large, perturbation theory is not available as an option and other techniques such guessing the exact answer or numerical simulations are needed.  The other thing to note is that $f_i \sim \partial_v^i f_0$.

Let us finish by finding the self consistent electric field and in particular its frequency.  To do this, we will use Maxwell's equation
\bea
\frac{d \vec E}{dt} &=& - \vec J \\
- E \omega \sin (\omega t )  &=& e \int dv \vec v f = e \int dv \, v \sum_i f_i \, .
\eea
When calculating the current, something nice happens.  Since $f_i \sim \partial_v^i f_0$, we see that
\bea
\int dv \, v \partial_v^i f_0 \sim \int dv \, \partial_v^{i-1} f_0 \sim  \partial_v^{i-2} f_0 \mid_{v = - \infty}^{v = \infty} = 0 \quad \forall  \, \, i \geq 2 \nn
\eea
Since the initial plasma is at rest, the only non-zero contribution to the current comes from $f_1$.  Something similar occurs when calculating the total number density, where only $f_0$ contributes.  We thus find that
\bea
J = -e \int dv v f_1 = \frac{e^2 n_e E}{m_e \omega} \sin \left ( \omega t \right ) \, .
\eea
Combining this with Maxwell's equation gives
\bea
\omega^2 = \frac{e^2 n_e}{m_e} \equiv \omega_p^2 \, .
\eea
We find that an oscillating plasma wave with zero wave-number, always has frequency equal to the plasma mass, regardless of the magnitude of the electric field.  Nonlinearities will necessarily involve $k\ne0$ modes.  If protons are added, one finds the expression $\omega_p^2 = e^2 n_e/m_e + e^2 n_p/m_p $.  There is no additional non-relativistic correction at any order in perturbation theory.

Finally, it is interesting to calculate the energy density in Langmuir waves.  The energy exists both in the kinetic energy of the electrons and the electric field giving
\bea
\rho = \frac{1}{2} E^2 \cos^2 \left ( \omega t \right ) + \frac{1}{2} m \int dv \, f (v^2 - \vth^2) = \frac{1}{2} E^2 \, . \nn
\eea
Much like how the density and current only get contributions from $f_0$ and $f_1$, the kinetic energy only gets non-zero contributions from $f_2$.

\subsection{$k \ne 0$ Instabilities}

At first glance, one's first intuition is that when studying a $k=0$ applied force, that one would only need to keep track of $k=0$ physics and that all $k \ne 0$ modes can be neglected.  In this section, we seek to impart that in plasma physics one can almost never neglect the $k\ne 0$ modes.  One particular $k$ mode of interest is the inverse Debye length
\bea
k \sim \frac{1}{\lambda_D} \equiv \frac{\omega_p}{\vth} \,.
\eea
The Debye length is the screening length of a plasma, and the intuition in that any wavelength smaller than the Debye length will be quickly screened.  This intuition is sound, as typically all wavelengths smaller than the Debye length are Landau damped while modes longer than the Debye length may or may not be unstable to grow.

Instabilities are most easy to find by studying the dispersion relationships and finding imaginary frequencies.  The dispersion relation of plasma waves are determined by the dielectric function defined as $\epsilon(\omega,k) \div E = \rho$.  Zeros of the dielectric function determine the dispersion relationship with a simple example dielectric function being $\epsilon = 1 - \omega_p^2/(w^2-k^2)$.

To get the dispersion relationship we solve Maxwell's and Vlasov's equations to first order in perturbations
\bea
f &=& f_0 + \df \\
\partial_t \df + i k v \df &=& \frac{q}{m} E \partial_v f_0 \nn \\
E &=& \frac{i}{k} q \int dv \df \, . \nn
\eea
We will solve these equations using the Laplace transform as it allows one to solve the evolution of perturbations as an initial value problem, as was originally done by Landau.
\bea
\df ( p ) = \int_0^\infty dt \, e^{-p t} \df (t) \, .
\eea
Turning the crank and solving for the dielectric constant, we obtain
\bea
\epsilon(p,k) = 1 - \sum_\alpha \frac{\omega_{p}^{\alpha,2}}{k^2} \frac{1}{n_\alpha} \int dv \frac{\partial_v f_0}{v - i p/k} \, .
\eea
The contour integral is taken along the real axis.  If there are poles on or below the real axis, the contour is deformed to extend below them.  This amounts to picking up the residue of any pole below the real axis and half the residue of any pole on the real axis.

We can find any instability by solving $\epsilon = 0$ with $p = - i \omega + \gamma$.  In the $\gamma \ll \omega, k \, \vth$ limit, we can Taylor series in $\gamma$ and solve to get the dispersion relationship and the damping or growth
\bea
\label{Eq: dis-gamma}
\text{Re} \, \epsilon(-i \omega,k) = 0 &\qquad& \text{Dispersion} \\
\gamma = - \frac{\text{Im} \, \epsilon(-i \omega,k) }{ \partial_\omega \text{Re} \, \epsilon(-i \omega,k)}  &\qquad& \text{Damping or Growth} \nn \, .
\eea
The real and imaginary parts of the dielectric function can be written as
\bea 
\text{Re} \, \epsilon(-i \omega,k) &=& 1 - \sum_\alpha \frac{\omega_{p}^{\alpha,2}}{k^2} \frac{1}{n_\alpha} \mathcal{P} \int dv  \frac{\partial_v f_0}{v-w/k} \nn \\
\text{Im} \, \epsilon(-i \omega,k) &=& - \sum_\alpha \frac{\omega_{p}^{\alpha,2}}{k^2} \frac{\pi}{n_\alpha} \partial_v f_{0 \alpha} (v = \frac{\omega}{k}) \, , \label{Eq: epsilon}
\eea
where $\mathcal{P}$ indicates the principle value of an integral.

While Eq.~\ref{Eq: dis-gamma} and Eq.~\ref{Eq: epsilon} can be used to calculate any instability, the intuition for them comes from following what happens to particles when they are hit by a wave.  Only particles traveling near the speed of the wave are relevant, as they are the only ones who are significantly affected by the wave.  Particles moving faster than the wave are slowed down while particles moving slower than the wave are sped up.  If there are more particles moving slower (faster) than the wave, the derivative is negative (positive), then the bath has a net gain (loss) of energy.  The  gained (missing) energy comes from the wave, damping (growing) it.

\subsubsection{Landau Damping of Langmuir Waves}

Langmuir waves are waves in the limit there the phase velocity of the wave is much faster than the thermal speed, $w/k \gg \vth$.  Turning the crank on Eq.~\ref{Eq: dis-gamma} gives
\bea
\omega^2 &=& \omega_p^2  + 3 k^2 \vth^2 + \cdots \\
\gamma &=& \frac{\pi}{2} \frac{\omega^3}{k^2} \frac{1}{n_e} \partial_v f _0(\omega/k) \, .
\eea
The expression for $\gamma$ indicates that for $f_0$ that are monotonically decreasing, that there is only ever damping.  There are instabilities if the slope of $f_0$ is ever positive, i.e. if there is ever a local minimum for $f_0$.

\subsubsection{Ion - Acoustic Waves}

Ion-acoustic waves are an example of a (sound) wave that is not necessarily strongly damped.  For reasons that will become clear in a moment we will take the electrons to be hotter than the ions and moving with a bulk velocity $u_e \ll \vthe$.  Long lived ion-acoustic waves occur when 
\bea
\vthe \gg \frac{w}{k} \gg \vthi \,.
\eea
Taking thermal distributions and using Eq.~\ref{Eq: dis-gamma}, we find the dispersion relation
\bea
\omega^2 &=&  \frac{c_s^2 k^2}{ 1 + \frac{k^2 {\vthe}^2}{\omega_{p,e}^2}} \nn \\
c_s^2 &=& \frac{q_i T_e}{e m_i} \, . \nn
\eea
In this wave, the hot electrons provide the pressure that moves the wave, while the ions provide the inertia.  This is why the sound speed of the wave depends on the temperature of the electrons but the mass of the ions when $T_e \gg T_i$.

Damping and growth of ion acoustic waves can be found using Eq.~\ref{Eq: dis-gamma} to give
\bea \label{Eq: ion growth}
\gamma = - \sqrt{\frac{\pi}{8}} \frac{\omega^3}{k^2 {\vthe}^3} \frac{e m_i}{q_i m_e} (\frac{\omega}{k} - u_e).
\eea
For small k modes, $\omega/k = c_s$ and we get the interpretation that if the bulk flow of the electrons is subsonic, then the ion acoustic wave is damped.  If the bulk flow of the electrons is supersonic, then the ion acoustic wave has an exponential growth.  This indicates that large k modes acoustic waves are generated if any external force drives the electrons to move supersonically.

In the $u_e \gg c_s$ limit, the k mode that grows the fastest is
\bea \label{Eq: kmax}
k = \frac{\omega_{p}^e}{\sqrt{2} \vthe} \, .
\eea
The general expectation is that $k \sim 1/\lambda_{D}^e$ is often the mode that grows the fastest.  Higher $k$ modes are damped by Debye screening.

The $T_e \gg T_i$ approximation is not always valid, e.g. it is not true before the dark photon heats the plasma.  All is not lost, as the dispersion relationship can be written in general as
\bea
Z(\xi) &\equiv& \frac{1}{\sqrt{\pi}} \mathcal{P} \int_{-\infty}^{\infty} du \frac{e^{-u^2}}{u-\xi} \\
k^2 \frac{{\vthe}^2}{2 \omega_{p,e}^2} &=& Z'(\frac{\omega/k - u_e}{\sqrt{2} {\vthe}}) + \frac{q_i T_e}{e T_i} Z'(\frac{\omega/k}{\sqrt{2} \vthi}) \, .\nn
\eea
Armed with this dispersion relationship, we can now use Eq.~\ref{Eq: dis-gamma} and Eq.~\ref{Eq: epsilon} even when $T_e = T_i$.

\subsection{Driven Langmuir Waves}

Let us now ask what happens when there is a small applied electric field, whose frequency exactly matches the plasma frequency.  In particular, we will apply an electric field
\bea
E_{\rm applied} = \epsilon E' \cos \left ( \omega_p t \right )
\eea
and see what happens.  Of course, this is motivated by the effect of dark photon dark matter.

\subsubsection{$k = 0$}
We will first calculate what occurs to the $k=0$ mode.  We begin with the Vlasov equation and Maxwell's Equation
\bea
\partial_t f &=& \frac{e}{m_e} \left ( E(t) + \epsilon  E' \cos \left ( \omega_p t \right ) \right ) \partial_v f \\
\partial_t E &=& - J = e \int dv \,  v f \, .
\eea
Taking the time derivative of Maxwell's equation and using the Vlasov equation, we find
\bea \label{Eq: k0growth}
\partial_t^2 E &=& e \int dv \, v \partial_t f \\
&=& - \omega_p^2 \left ( E(t) + \epsilon  E'  \cos \left ( \omega_p t \right ) \right ) \, , \nn
\eea
where we have integrated by parts and used $\int dv f = n_e$ to obtain the final expression.  Solving this equation, we find that
\bea \label{Eq: Ek=0}
E(t) = - \frac{1}{2} t \omega_p \epsilon  E'  \sin \left ( \omega_p t \right ) 
\eea
The electric field (energy) in the Langmuir waves grows as $t$ ($t^2$) when it is driven at exactly the resonant frequency.
This growth continues until nonlinear effects from the $k \ne 0$ modes generated by parametric resonance, as discussed in the next subsection, change the resonant frequency of the $k = 0$ mode and the pumping goes off-resonance and energy is no longer injected into the system.

\subsubsection{$k \ne 0$}

In this section, we discuss how $k \ne 0 $ modes are generated by a combination of the ponderomotive force and parametric resonance.
Our estimates will be a bit heuristic as the full calculation is tedious, non-illuminating, and too complicated for this appendix.  Motivated readers can read through Ref.~\cite{1972JETP...35..908Z} for a more rigorous derivation.

The starting point is the ponderomotive force.  An extremely fast review of the ponderomotive force is that in the presence of a fast oscillating but slowly changing in distance electric field $E = E(x) \cos \left ( \omega t \right )$, a particle moves as
\bea
\ddot x = \frac{e E(x)}{m} \cos \left ( \omega t \right ) \, .
\eea
$x(t)$ around a location can be expanded into a fast oscillating mode $x_1$ and a slow oscillating mode $x_2$. The fast oscillation around $x_2$ can be found to be $x_1 = - \frac{e E(x_2)}{m \omega^2} \cos \left ( \omega t \right )$ and the slow oscillation
\bea
x_2 \approx -\frac{e^2}{4 m^2 \omega^2} \nabla E^2(x_2) \,,
\eea
where the $\approx$ comes from a Taylor expansion. We see that the particles are moving in a potential
\bea
\Phi_p= \frac{e^2 E^2}{4 m \omega^2} \,,
\eea
the ponderomotive potential. In the adiabatic approximation, this means that the density perturbations induced by the electric field of non-zero $k$ modes is
\bea
n &=& n_0 e^{- \frac{e^2 E^2}{4 \omega^2 T}} \nn \\
\delta n &\approx& - \frac{E^2}{4 T} \approx - \frac{E_{k=0} \delta E}{2 T}\, .
\eea
When writing $\delta n$, we took $\omega = \omega_p$ as this is the case we are interested in.  We will be interested in perturbing around the solution shown in Eq.~\ref{Eq: Ek=0} and finding an instability.  As such, we have also expanded $E^2$ to leading order in $\delta E$.  The heuristic nature of this derivation appears as the $E_{k=0}^2$ piece is not present, as a uniform electric field cannot change the number density.  Namely, the assumption that the number density can move particles across a distance $1/k$ to change the local number density fails when $k=0$.

Finally, we look at how density perturbations change the plasma mass as
\bea
\delta \omega_p^2 = \frac{e^2 \delta n}{m_e} = - \frac{e^2 E_{k=0} \delta E}{2 m_e T} \, .
\eea

We can now expand Eq.~\ref{Eq: k0growth} around the solution Eq.~\ref{Eq: Ek=0} to study perturbations, giving
\bea \label{Eq: guess}
\left ( \partial_t^2 - 3 {\vthe}^2 \nabla^2 + \omega_p^2 \right ) \delta E = \frac{e^2 E_{k=0}^2}{2 m_e T}  \delta E \\
= \frac{e^2 \epsilon^2  E'^2 \omega_p^2 t^2}{16 m_e T} \left ( 1 - \cos \left ( 2 \omega_p t \right )\right )  \delta E \nn \, .
\eea
If we are in the regime where $\epsilon$ is very small so that the growth of the magnitude of the electric field is a small effect, then this is a Mathieu function
\bea
\partial_\tau^2 E + \left ( a - 2 q \cos \left( 2 \tau \right ) \right ) E = 0 \, ,
\eea
where $\tau = \omega_p t$, $a = 1 + 3 v_{\rm th}^{e,2} k^2/\omega_p^2$, and $q \sim v_q(t)^2/v_{\rm th}^{e,2}$.
As is well known, the Mathieu equation has instabilities.  As $q \ll 1$, the instability band that grows the fastest corresponds to
\bea
a = 1 \pm q \qquad \gamma = \frac{q}{2} \qquad E \sim e^{\gamma \omega t} \, .
\eea
This means that the low k modes with $k \lambda_{D}^e \lesssim v_q(t)/\vthe$ experience exponential growth with a rate of $\omega_p v_q^2(t)/v_{\rm th}^{e,2}$.  Of course the full equations are horrendously nonlinear, so this growth does not remain exponential for long. 

\section{Simulation details}
\label{app:sim}

The particle-in-cell (PIC) simulations reported in this letter were performed using the SHARP code \cite{sharp,sharp2}. We use the code to solve the Vlasov-Maxwell equations (Eq.~\eqref{Eq: vlasov-maxwell}) in 1+1D, meaning that only the $(x,v_x)$ subspace of the phase space is used to describe the dynamics of the particle distribution function $f_\alpha$ for both electrons and ions. The total momentum in the simulation is conserved exactly, and energy conservation is well-controlled due to the use of fifth-order spline functions in both the deposition (of charge density and currents) and the back-interpolation step (which computes the Lorentz force on particles). Particle velocity updates are performed using the Vay algorithm \citep{Vay-2008}. The code has been extensively utilized to study various phenomena, including beam-plasma instabilities \citep{resolution-paper,sim_inho_18,th_inho_20}, cosmic-ray-driven instabilities \citep{sharp2,Lemmerz+2023,Lemmerz2025}, and the formation of shocks in electron-ion plasmas \citep{Shalaby+2022ApJ,Shalaby2024ApJL,Shalaby2025ApJ}.

In all simulations, we self-consistently evolve both ion and electron dynamics. Both species are initialized with a uniform spatial distribution across the computational domain and use a realistic ion-to-electron mass ratio $m_i/m_e = 1836$. The initial velocity distribution for both species is a Maxwellian (Gaussian) given by
\bea f_\alpha(v_x) =  \frac{n_\alpha}{ \sqrt{2 \pi k_B T_\alpha/m_\alpha}} e^{- \frac{ m_\alpha v_x^2}{2 k_B T_\alpha}} .
\label{eq:dist}
\eea 
Where, $n_\alpha$ is the number density, $k_B T_e = k_B T_i = 10^{-3} m_e c^2$. This corresponds to electron and ion Debye lengths of $\lambda_D \sim 0.033 c/\omega_p$. The cell size used in all simulations is $\Delta x = 0.04 c/\omega_p$, meaning the initial Debye length is marginally resolved. While waves below this scale will be strongly Landau damped, this choice of cell size helps mitigate numerical heating \cite{birdsall+1980}.

In all simulations, we use $1000$ computational cells, corresponding to a domain size of $L = 40 c/\omega_p$. We have verified that using larger box sizes yields the same quantitative results as those presented in this letter.

We note here that particle distribution in Eq.~\ref{eq:dist}, the thermal speed $ v_{\rm th} $ is defined such that
\bea 
v^2_{\rm th}  &\equiv&  \langle (v_x - \langle v_x \rangle )^2\rangle = \int_{-\infty}^{\infty} dv_x  (v_x - \langle v_x \rangle )^2 f(v_x) \nonumber \\ 
&=& \frac{ k_B T}{m}
\eea
The initial average velocity in the $x$-direction is zero, i.e., 
\( \langle v_x \rangle \equiv \int_{-\infty}^{\infty} dv_x \, v_x f(v_x) = 0 \).
We define the thermal energy density of each species (electrons or ions) as follows.\footnote{If we were to evolve all velocities, i.e., include the dynamics in $v_y$-$v_z$ phase space, the temperature would be the sum of the temperatures in the three velocity directions. Consequently, the square of the thermal speed would be the sum of the averages of the \emph{squared} velocities in each direction.} 
\bea
E_{\rm th}^\alpha  = \frac{n_s m_s  {v_{\rm th}^\alpha}^2 }{2} =  \frac{n_\alpha k_B T_\alpha}{2}
\label{eq:Eth}
\eea
That is, in all simulation we thus fix $v_\mathrm{th}^e = 3.1 \times 10^{-2} c$ to facilitate direct comparison between different simulations.
The change in this thermal energy is what is plotted in Fig.~\ref{fig:Efield} and Fig.~\ref{fig:LZcase}.

In our kinetic simulations, the electrons and ions are represented by macroparticles, where each macroparticle represents a collection of real particles. Consequently, the Poisson noise (or shot noise) in the simulation, which arises from the finite number of macroparticles, is typically much greater than that in the actual plasma whose evolution we are investigating.
The contribution of this Poisson noise to the electric field and its spectrum is computed analytically in Appendix D of \citep{sharp}. The specific contribution to the $k=0$ mode is given by
\bea
\epsilon_{\mathrm{noise}}  = \frac{L^2}{12 N_p N_x} n_e m_e c^2,
\label{eq:PE_noise}
\eea
Here, $L$ is the domain size in units of the electron skin depth $c/\omega_p$, $N_p$ is the total number of macroparticles, and $N_x$ is the number of computational cells. Compared to the analytical form in Appendix D of \citep{sharp}, we divide by $N_x$ to compute the average contribution per computational cell.

\begin{figure}
    \centering
    \includegraphics[width=1\linewidth]{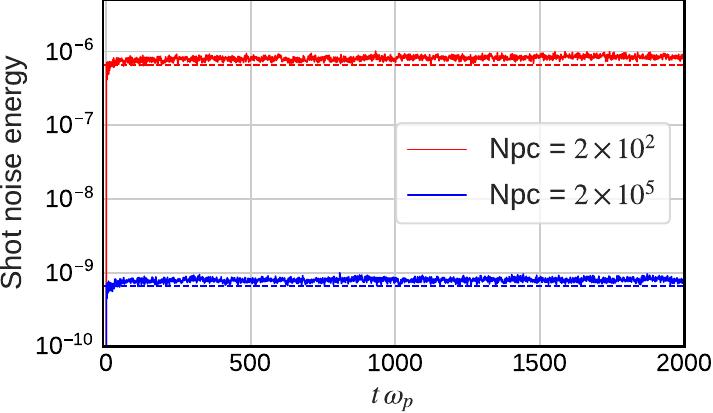}
    \caption{\label{fig:A00}
    Evolution of the shot noise energy density (in units of $n_e m_e c^2$; solid lines) in undriven ($A_0 = 0$) PIC simulations with different numbers of particles per cell. The analytically expected shot noise level, computed using Eq.~\eqref{eq:PE_noise}, is shown with dashed lines. We note that for the typical parameters used in our simulations, the thermal energy density (in units of $n_e m_e$) given by Eq.~\eqref{eq:Eth} is $5 \times 10^{-4}$, which is much higher than the noise level in all presented simulations.
    }
\end{figure}

To validate the shot noise calculation, Fig.~\ref{fig:A00} shows the evolution of the electric potential energy (per cell) in two undriven simulations ($A_0 = 0$) with $N_x = 1000$ and grid spacing $h = \Delta x ~\omega_p/c = 0.04$. We plot the evolution (solid lines) for simulations with an average number of particles per cell of $N_\mathrm{pc} = 2 \times 10^2$ and $2 \times 10^5$. The expected shot noise level, computed using Eq.~\eqref{eq:PE_noise}, is shown with dashed lines. We observe excellent agreement between the theoretical prediction and the measured noise levels in the simulations.
We also note here that for such typical numbers chosen in our simulations, the thermal energy density (in units of $n_e m_e c^2$) given by Eq.~\eqref{eq:Eth}, is $5 \times 10^{-4}$, much higher compared to the noise in all simulations.

For resonant driving, the total driven potential energy (per computational cell) grows with time, using Eq.~\ref{Eq: Ek=0}, as follows 
\begin{align*}
\epsilon_{\mathrm{driven}}  =  \frac{A_0^2}{8} ( \omega_p t)^2 n_e m_e c^2.
\end{align*}
Where, $ A_0 \equiv \epsilon  E'/\sqrt{n_e m_e c^2/\epsilon_0} = \epsilon e E'/m_e \omega_p c = v_q^{\rm D}/c$, and $\epsilon_0$ is the vacuum permittivity.
Thus, the time required for this driven energy to exceed the initial Poisson (shot) noise energy is found as follows
\bea
\epsilon_{\mathrm{driven}} > \epsilon_{\mathrm{noise}}
\Rightarrow 
 \omega_p t 
 > \Bar{t}_\mathrm{noise} \equiv \frac{L}{A_0 \sqrt{3 N_p N_x/2}}.~~~~
\eea
In our simulations, we mitigate this noise by initializing all particles equally spaced, with ions and electrons colocated throughout the computational domain. This configuration ensures that initially $\epsilon_{\mathrm{noise}} = 0$. For such a setup, it takes approximately $t \omega_p \sim 3\text{--}4$ for the driven energy to be converted into de-coherent (high-$k$) noise. Therefore, in all our resonant conversion simulations, we increase the total number of particles for lower values of $A_0$ such that the characteristic noise onset time satisfies $\bar{t}_{\mathrm{noise}}  \sim 2.4$.
On the other hand, in the Landau-Zener simulations, the driven potential energy does not grow initially. Consequently, for these runs we increase the number of particles such that $\bar{t}_{\mathrm{noise}} \leq 0.02$.
That is, the average (initially fixed) driven energy density is $50$ times larger than that of the shot noise.

The impact of the dark photon on the electron-ion plasma is incorporated as an external electric field with a uniform amplitude that evolves over time \(A_0 \cos(\omega t)\) across all computational cells. This configuration represents a \(k=0\) external field.
For resonance simulations, the frequency is fixed at \(\omega = \omega_p\). In Landau-Zener simulations, the frequency evolves as \(\omega(t) = 0.8\omega_p + 0.1\omega_p t/5000\).

\section{Collection of numerical results}\label{app:simresult} 

In this appendix, we characterize the rate at which energy is injected from the dark photon dark matter driver into the Standard Model electron-ion plasma, and then present the long-term behavior of various simulations in the main body of this letter.

\begin{figure}
\includegraphics[width=1\linewidth]{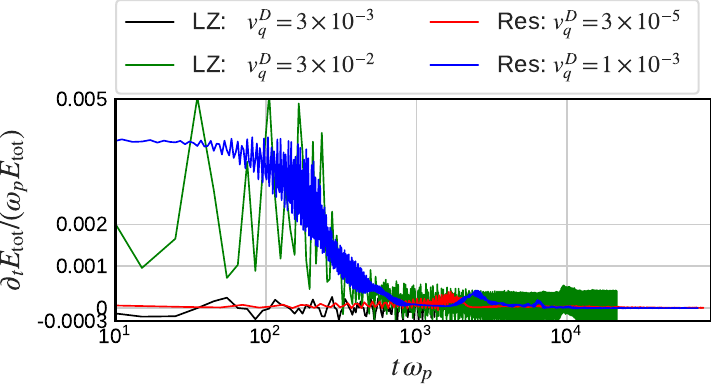}
\caption{\label{fig:rate}
Rate of energy injection into the electron-ion plasma in various simulations. Res: for resonant conversion simulations, and LZ: for Landau-Zener simulations.
}
\end{figure}

We define (fractional) energy injection rate into the electron-ion plasma as:
\bea
\mathrm{rate} = \frac{\partial_t E_\mathrm{tot}}{\omega_p E_\mathrm{tot}}
\label{eq:rate}
\eea
where $E_\mathrm{tot}$ is the total energy in the electron and ion species. 
Figure~\ref{fig:rate} shows this rate for various simulations, including both resonant and Landau-Zener cases.
Simulations with strong driving show higher energy injection rates (blue and green curves) compared to those with weaker driving (red and black curves).
For comparable driving strengths (black and blue curves), the energy injection in the resonant simulation---where at $t=0$ the DPDM oscillates at exactly the plasma frequency---is much higher than in the Landau-Zener case, where the DPDM oscillates at $80\%$ of the plasma frequency at $t=0$ and $120\%$ of the plasma frequency at $t=10^4$ .

For Landau-Zener simulations, Figure~\ref{fig:rate} shows that after a tiny initial injection of energy early on, the energy injection rate drops to roughly zero after nonlinearity becomes important when 
\begin{equation}
    \frac{1}{1-\omega_p^2/m_{A'}^2}\frac{\varepsilon E'}{m_e\omega_p} \simeq \vthe.
\end{equation}
The subsequent injection of energy is suppressed, with the most dramatic effect being that energy transfer during the expected strong resonance around $ \omega_pt = 5000$ completely disappears.

\begin{figure}
    \includegraphics[width=1\linewidth]{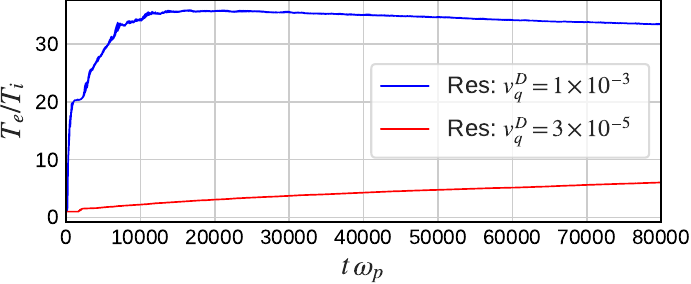}
    \caption{\label{fig:TeTi}
    Long-term evolution of the electron-to-ion temperature ratio in resonant conversion simulations. }
\end{figure}

In Fig.~\ref{fig:TeTi}, we show the temperature ratio $T_e/T_i$ for the two resonant conversion simulations~\footnote{We cannot effectively distinguish between energy that is already fully thermalized and energy stored in the high $k$ modes of oscillations of the electrons and ions; as a result, the temperatures presented here shall be understood as a measure of the total kinetic energy of the electrons and ions (with the $k=0$ component removed).}.
For the weaker driving case (red line), this temperature ratio is still increasing with time as a result of slow electron heating towards the end of the simulation. On the other hand, in the simulation with stronger driving (blue line), this ratio very quickly exceeds 30 and stops increasing. At this point, as we explained in the main text, the final state is stable with $T_e/T_i \gg 1$. In both simulations, the resonant conversion is suppressed when the temperature ratio is only slowly changing, and we expect the resonant conversion to remain suppressed on time scales much longer than our simulation time. This can also be seen in the animation for various simulations that can be found in \cite{Video_DPDM}.

\begin{figure*}
    \centering
    \includegraphics[height=5.65cm]{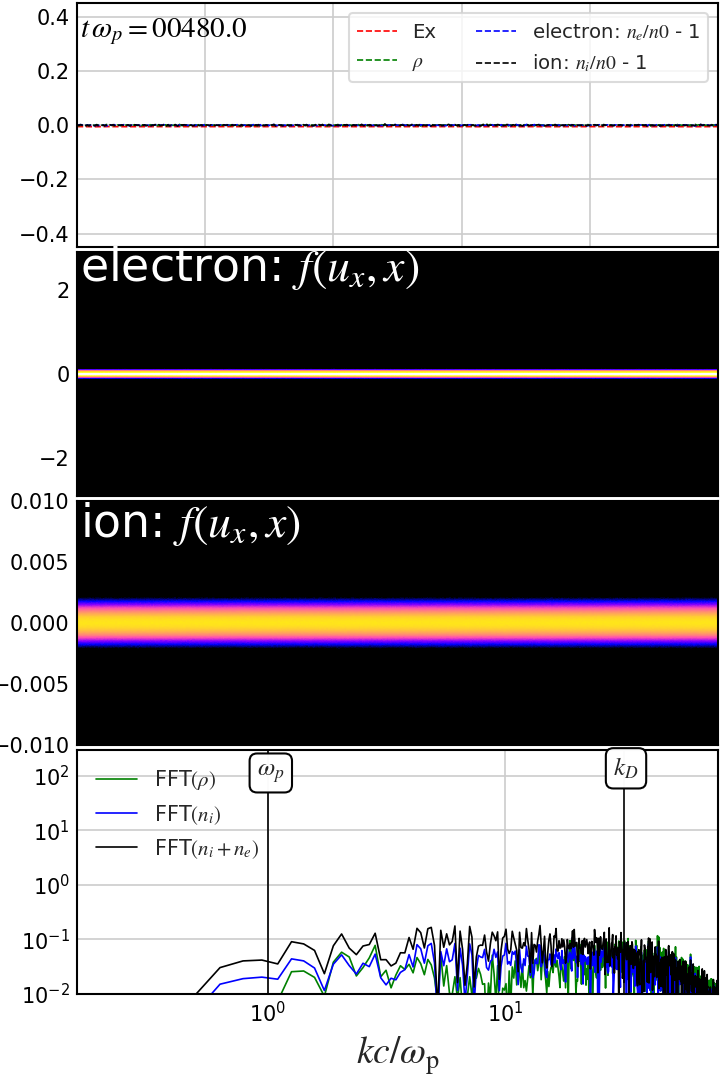}
    \includegraphics[height=5.65cm]{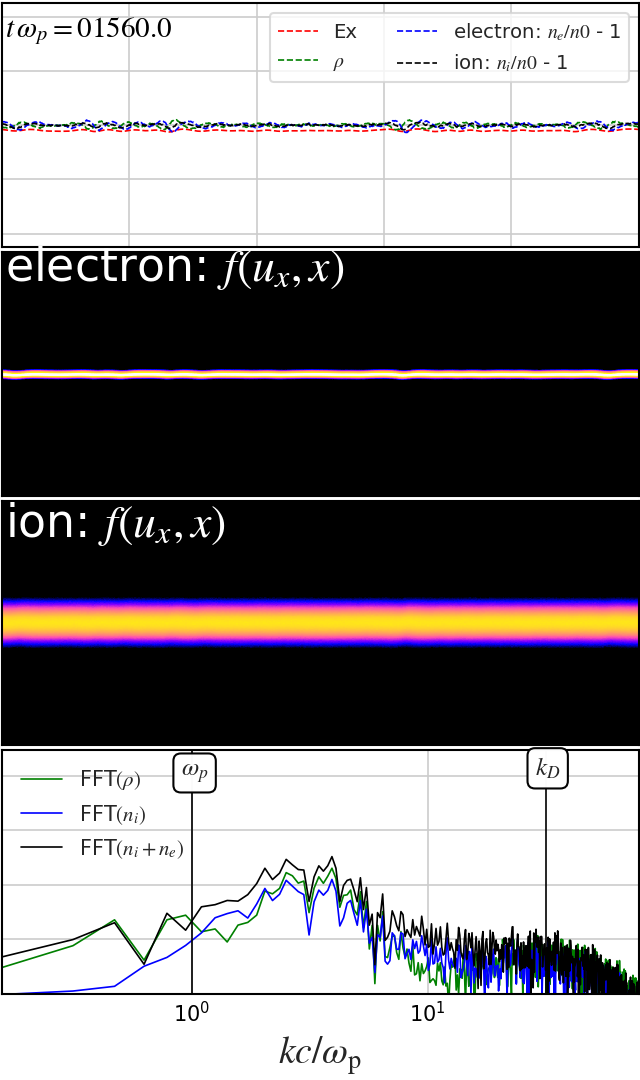}
    \includegraphics[height=5.65cm]{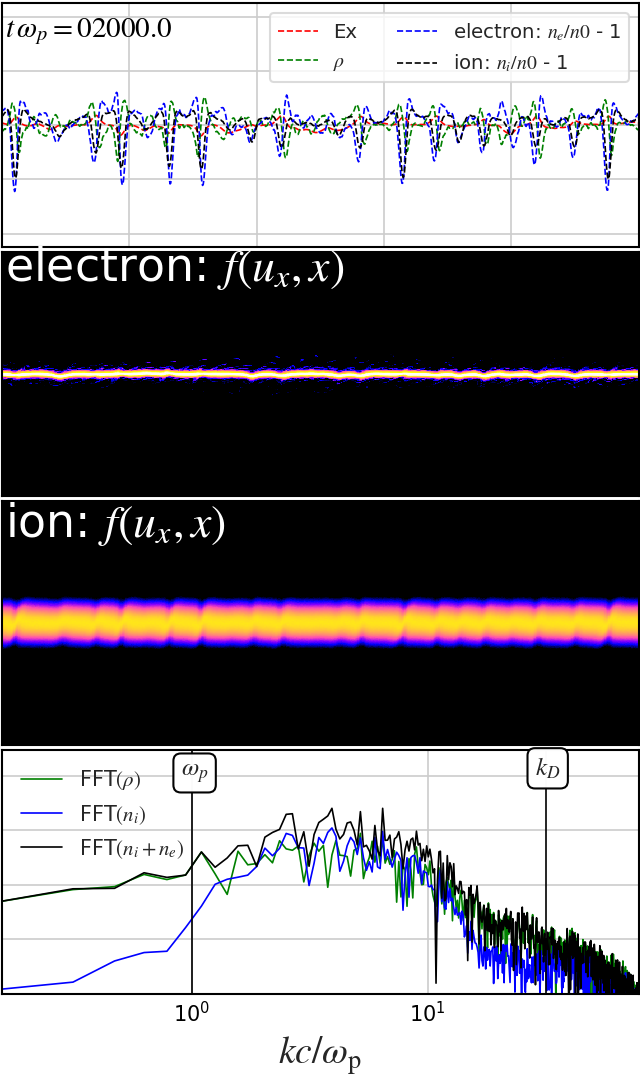}
    \includegraphics[height=5.65cm]{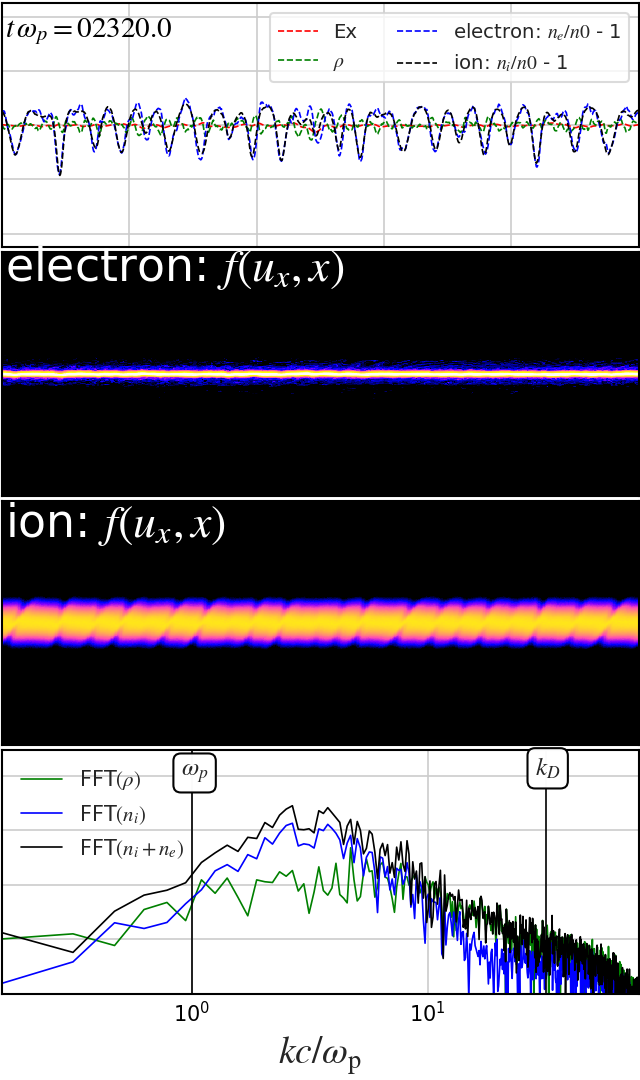}
    \includegraphics[height=5.65cm]{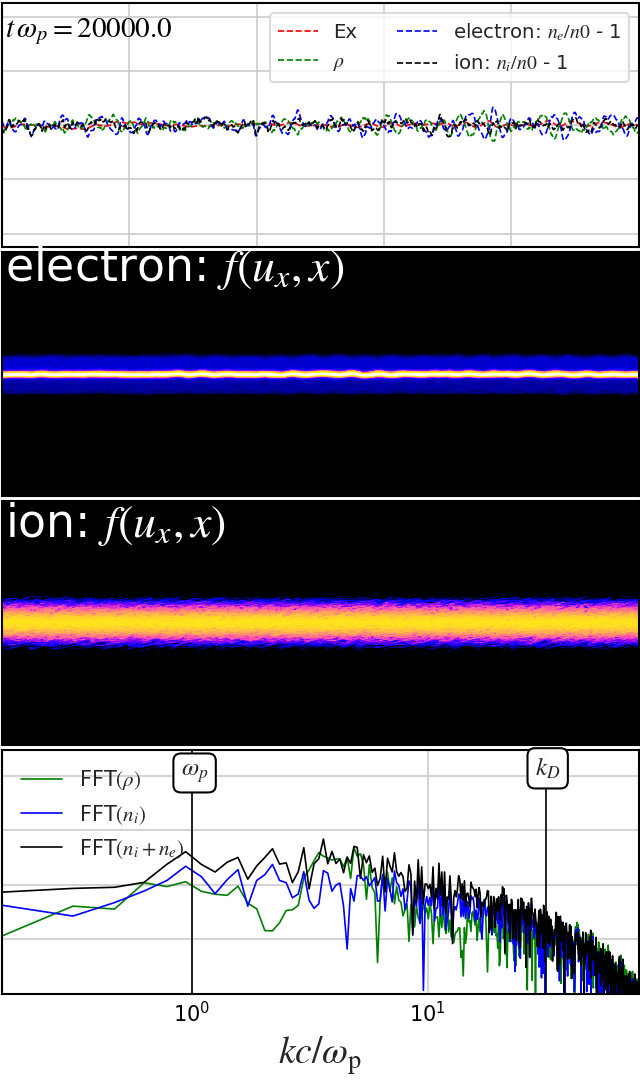}
    \caption{Five time frames of the evolution of the {\it slow growth} ($v_q^D/\vthe = 10^{-3}$) simulation~\cite{Video_DPDM} at $\omega_p t = 480, 1560, 2000, 2320$ and $20000$ from left to right. The upper three panels are in real space, while the bottom panel is in $k$ space. See text in App.~\ref{app:simresult} for detailed descriptions. Here, $n_e \, (n_i)$ is the electron (ion) number density, and $\rho$ is the charge density.}
    \label{fig:panels}
\end{figure*}
In the remainder of this appendix, we will provide a guide to understanding the videos shown in~\cite{Video_DPDM} by mainly providing (in Fig.~\ref{fig:panels}) and describing some of the important time in the evolution of the system driven by DPDM with the example of $v_q^D/\vthe = 10^{-3}$ simulation (lower panel of Fig.~\ref{fig:Efield}). 

First, for $\omega_p t \lesssim 480$, the system contains mostly $k= 0$ mode of the Langmuir wave, which shows up as an oscillating electric field (red line) and charge density $\rho$ (green line) in the first panel. The visible oscillation is almost completely spatially independent. At the same time, the velocity of the electron $u^e_x/c$ oscillates (second panel) while that of the ion $u^I_x/c$ does not show any visible oscillation (third panel)\footnote{The new notation $u_x$ shows up in simulation output, and is equal to $\gamma v_x$. In all of the simulations we presented, the electrons and ions are not relativistic enough for this difference to be visible.}. In the bottom panel, it can be seen that the $k\neq 0$ mode already starts growing.

At around $\omega_p t = 1560$, growth of $k\neq 0$ modes becomes visible in not only the bottom panel, but also the upper two panels.
Both $k\neq 0$ modes of the Langmuir wave and the ion acoustic wave grows. Such a growth period terminates at around $\omega_p t = 2000$, when it is clear that the ion density shown as black line in the top panel (as well as total density) fluctuations become significant. Similarly, one can see clear spatial variation of the total density in all other simulations when the growth of the electric field energy terminates. 

At around $\omega_p t = 2320$, another feature worth noting emerges. Whereas the neutral density fluctuations remains ( black lines in top or bottom panel), the charge density fluctuations (green lines in top and bottom panel) disappears. This corresponds to Fig.~\ref{fig:Efield} as the time when the red solid and dashed line reaches the minimum. At this point, the resonance is completely lost and the energy stops transferring from the DPDM to the $k=0$ Langmuir wave. The same behavior also occurs for the $v_q^D/\vthe = 0.03$ simulation around $\omega_p t = 800$, though due to the significant overshooting, both $u^e_x/c$ and $u^I_x/c$ have larger oscillations and the system is heated significantly at this point.

The evolution afterwards lead to populating the system with a variety of $k$ modes of both ion acoustic and Langmuir waves up to the inverse of the Debye length $k_D$ (bottom panel) and the system evolves towards thermalization. However, it is clear that even at the end of the simulation, the system still hosts significant ion acoustic and Langmuir wave oscillations at $k\neq 0$.

\section{Dark Photon Dark Matter Constraints}\label{app:constraints}

In this appendix, we present some more details about how we derive the updated cosmological constraints. Observationally, these cosmological constraints come from two qualitatively different considerations. Before recombination, photon and electrons thermalize faster than the Hubble time scale at the time, and the ionization fraction is unity. During this period, energy injected into the Standard Model plasma is mostly transferred to radiation eventually. For energy injected before $a \approx 5\times 10^{-7}$, the Standard Model thermal bath, including electrons, ions and photons, would fully thermalize. The injected energy raises the temperature of this bath compared to the already decoupled neutrinos. Consequently, this scenario is constrained by measurements of the effective number of neutrino species $N_{\rm eff}$, and the total energy injected into the thermal bath shall be smaller than $\mathcal{O}(10^{-2})$ of the total radiation energy density. For energy injected after $a \approx 5\times 10^{-7}$, the Standard Model bath cannot fully thermalize. This leaves $\mu$ and y-type spectral distortion as observable signal of energy injection in this period. The non-observation of spectral distortions in COBE/FIRAS data puts a constraint on the maximal energy injection at around $\Delta \rho_{\gamma}/\rho_{\gamma} \lesssim 10^{-4}$~\cite{Arias:2012az}. Such a spectral distortion limit applies to energy injection until $a \approx 10^{-3}$. Between the end of recombination and the beginning of reionization, the limits on energy injection mainly stem from measurement of the total optical depth to last scattering $\tau \approx 0.06$~\cite{Planck:2018vyg}. Whereas recent analyzes suggest that possibly $\tau$ is larger than reported by Planck~\cite{Sailer:2025lxj}, it is quite robust that the ionization fraction cannot increase to more than $10^{-2}$ during the dark ages~\cite{Cheng:2025cmb}. After reionization, constraints on energy injection can come from Helium reionization, as well as Lyman-$\alpha$ measurements~\cite{Trost:2024ciu,Liu:2020wqz,Caputo:2020bdy,McDermott:2019lch,Witte:2020rvb,Slatyer:2015kla}. In particular, Lyman-$\alpha$ measurements suggest that the baryon temperature cannot increase to more than about $10^4\,{\rm K}$ based on line broadening~\cite{McDermott:2019lch,Witte:2020rvb,Liu:2020wqz}. 

\begin{figure}
    \centering
    \includegraphics[width=0.9\linewidth]{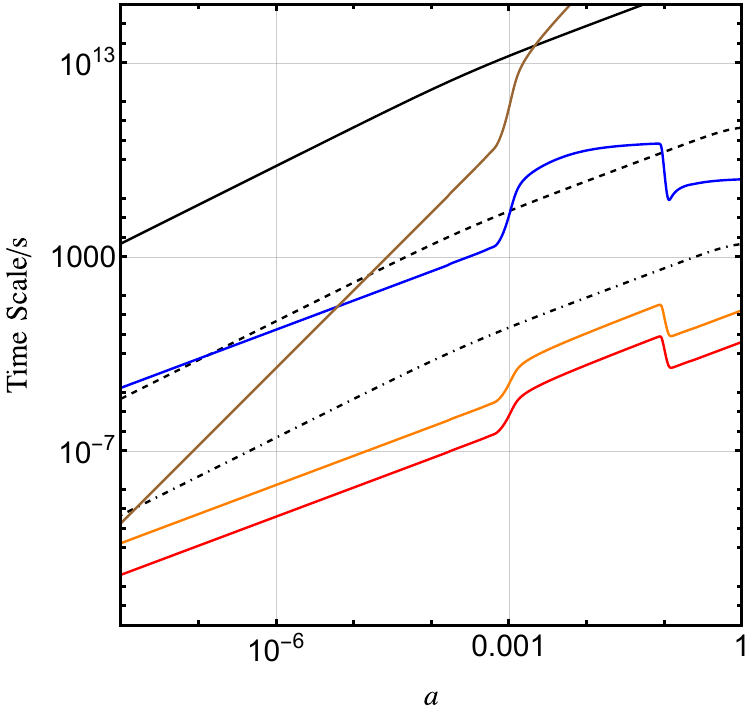}
    \caption{Various time scales of interest for estimating the limits of dark photon dark matter. Apart from the time scales described in Fig.~\ref{fig:velocity}, the black solid line shows the Hubble scale as a function of redshift, while the brown line shows the time scale of energy exchange between CMB photons and the electrons.}
    \label{fig:timescale2}
\end{figure}

\paragraph{Resonant conversion and a potential steady state} With dark photon dark matter, energy injection can both come from the burst of resonant conversion, as well as the longer term non-resonant conversion. As we established in the main text and App.~\ref{app:simresult}, in the absence of collisions, the system goes nonlinear after a very small amount of energy is injected into the electron ion plasma through resonant conversion. After the initial burst, the resonant conversion shuts off as the damping rate of ion acoustic wave onto ions becomes exponentially suppressed by the ratio $T_e/T_i$. This ratio, as we observe in the upper panel of Fig.~\ref{fig:Efield}, is fixed if a steady state were to be reached also during the Landau-Zener resonance. The collisions between electrons and ions ($10^{-6}\lesssim a \lesssim 10^{-3}$), as well as electrons with photons ($a \lesssim 10^{-5}$) tends to reduce this temperature ratio, and energy shall slowly be converted from DPDM to the Standard Model plasma to sustain this steady state for $\varepsilon \gtrsim 10^{-8}$ (see Fig.~\ref{fig:timescale2}).

\begin{figure}
    \centering
    \includegraphics[width=0.9\linewidth]{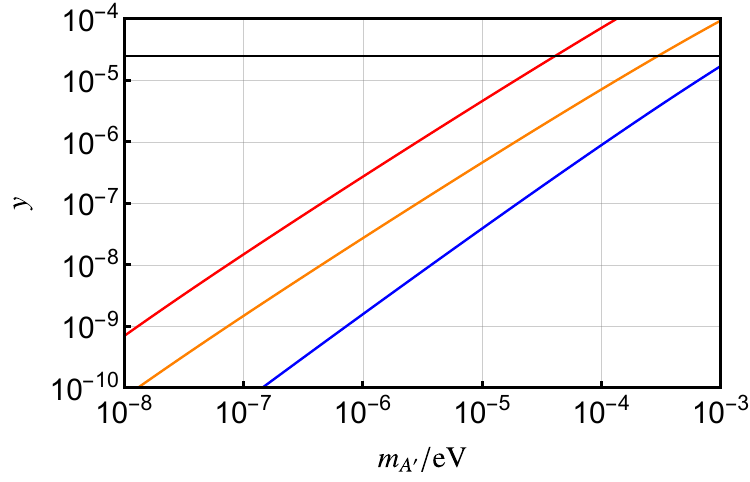}
    \caption{Possible sizes of $y$-distortion as a function of $\varepsilon$. The red and orange line corresponds to $\varepsilon = 10^{-7}$ and $10^{-8}$, respectively, while the blue line shows the $y$-distortion if the $T_e/T_{\gamma}$ can only be sustained during the period where stationary phase approximation is valid. The black solid shows the current limit on $y$-distortion ($2\times 10^{-5}$) as a comparison.}
    \label{fig:ydistortion}
\end{figure}

When the electron photon energy exchange rate dominates, energy density flows from electron to photon determined by the electron photon scattering rate of
\begin{equation}
    \nu_{e \gamma} \approx \frac{\pi^2 \sigma_T T_{\gamma}^4}{15 m_e}.
\end{equation}
Assuming that the ratio of $T_e/T_{\gamma} \approx T_e/T_{i} \approx \log [\omega_{p}^i/\nu_{e \gamma}] $, we can compute the amplitude of this $y$-distortion signal~\cite{Chluba:2011hw,Hill:2015tqa} as a function of $\varepsilon$ (shown in Fig.~\ref{fig:ydistortion}) directly with:
\begin{equation}
    y = \int {\rm d}t \frac{n_e \sigma_T (T_e-T_{\gamma})}{m_e} \propto \eta_B \varepsilon,
\end{equation}
where $\eta_B = 6.7\times 10^{-10}$ is the baryon to photon ratio, and $\sigma_T$ is the Thompson cross section. 
The linear scaling with $\varepsilon$ stems from the steady state that could have been reached in the early universe. In this estimate of the scaling of the $y$-distortion signal with model parameters $\eta_B \varepsilon$, we assumed that the temperature ratio can be sustained for the whole Landau-Zener transition time of $\varepsilon/H$, which, for the parameters of interest, is always larger than the duration where the stationary phase approximation is valid $\Delta t_{\rm sp}\approx ( m_{A'} H)^{-1/2}$. If the temperature ratio can only be sustained for a duration of $\Delta t_{\rm sp}$, then the $y$-distortion signal would be much weaker and $\varepsilon$-independent as long as $\varepsilon \gtrsim 10^{-9}$ (see blue line in Fig.~\ref{fig:ydistortion}). 
A large $T_e/T_{\gamma}$ ratio might not be sustainable in the very early universe, when the photon electron collision time approaches the ion plasma frequency $\omega_{p}^i$, which would weaken the constraints. We do not study this in more detail since the solar cooling constraints~\cite{An:2013yfc} are already much stronger than the cosmological constraints at those masses. 

For large enough $\varepsilon$, the time scale for the entire duration of a Landau-Zener resonant conversion can also be longer than the electron-ion energy exchange timescale. If the electron ion energy exchange timescale is also shorter than the electron photon energy exchange timescale, the temperature of the ions and electrons can continue to grow when a steady electron ion temperature is reached. In this case, the heating rate due to resonant conversion would be exponentially sensitive to $T_e/T_i$, the ratio that controls the damping of the ion acoustic wave, the perturbation that forbids resonant conversion from occurring. During this period, this ratio of $T_e/T_i$ can be fixed while both temperature increases. Such an increase could raise the temperature, roughly, until the ion electron energy exchange time scale $\tau_{ei}$, scaling as $T_e^{3/2}$, becomes longer than the duration of the resonance. Quantitatively, we can solve for this time growth and find that the electron can only increase by a factor of about 100 during this period. Such a difference does not show up visibly on the final constraint plot, as the non-resonant conversion we discuss later is more important.

\paragraph{Non-resonant conversion} A stronger constraint on the dark photon dark matter parameter space comes from non-resonant conversion, which slowly heats up the plasma on $1/H$ time scales. The heating rate depends on the collision rate between the electrons with ions as well as photons, and the total energy converted can be estimated to be
\begin{equation}
    \Delta \rho \approx \int \varepsilon^2   \rho_{\rm DM} \nu_{e}(T_e) \left ( \frac{m_{A'}^2}{\omega_p^2}\right )^{{\rm sign }[ \omega_p(t) - m_{A'}]} {\rm d} t, \label{eq:heatingnonres}
\end{equation}
where $\nu_{e}(T_e) = \nu_{e\gamma} + \nu_{ei}$ is the total collision rate of the electron. Depending on the relative size of the time scales $1/\nu_{e\gamma}$, $1/\nu_{ei}$ and $1/H$, there are three different regimes.

Firstly, for $a\lesssim 10^{-5}$, the electron-photon collision is the most frequent ($1/\nu_{e\gamma} < 1/\nu_{ei} < 1/H$). In this case, $T_e = T_{\gamma}$ and both shall not change substantially during heating for any $\varepsilon$ of interest. The total energy transfer can be estimated to be
\begin{equation}
    \frac{\Delta \rho}{\rho_{\gamma}} \approx \varepsilon^2 \frac{\rho_{\rm DM}}{\rho_{\gamma}} \frac{\nu_{e\gamma}}{H} |_{\omega_p = m_{A'}} \propto a^{-1},
\end{equation}
where the densities, as well as $ \nu_{e\gamma} $ and ${H}$ are functions of time. 

Secondly, for $10^{-5} \lesssim  a \lesssim 10^{-3}$, the electron-ion collision is the most frequent ($ 1/\nu_{ei} <1/\nu_{e\gamma} < 1/H$). In this case, in principle the temperature of the electrons and ions can grow as compared to the temperature of the photon. However, numerically, this temperature change is small for the smallest $\varepsilon$ that is constrained by this heating measurement, and as a result, Eq.~\ref{eq:heatingnonres} can again be simplified to be 
\begin{equation}
    \frac{\Delta \rho}{\rho_{\gamma}} \approx \varepsilon^2 \frac{\rho_{\rm DM}}{\rho_{\gamma}} \frac{\nu_{ei}}{H}|_{\omega_p = m_{A'}} \propto \varepsilon^2 a^{3/2} \propto \varepsilon^2 m_{A'}.
\end{equation}
Before recombination, the resulting limit is about $3000$ times weaker compared to the constraints reported before in~\cite{Arias:2012az}. The range of redshift when $X_e$ changes significantly, which interpolates between the orange and red shaded regions in Fig.~\ref{fig:money} requires more careful treatment, and we leave a dedicated study to future work.

\begin{figure}
    \centering
    \includegraphics[width=0.9\linewidth]{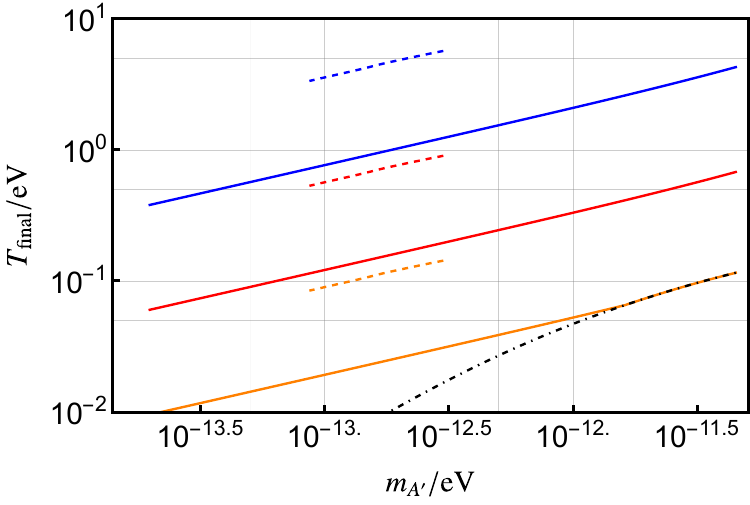}
    \caption{Terminal temperature for different $\varepsilon$ and $m_{A'}$ at the time period in the early universe when $m_{A'} \approx \omega_p$. The solid line is for the dark ages $20<z<500$, while dashed line for Lyman-$\alpha$ $2<z<6$. The blue, red and orange lines show the terminal temperature for $\varepsilon= 10^{-8}$, $10^{-9}$ and $10^{-10}$, respectively, while the black dot-dashed line shows the temperature of electrons without DPDM heating obtained from CLASS as a comparison.}
    \label{fig:terminalT}
\end{figure}

Thirdly, for $a\gtrsim 2\times 10^{-3}$, $ 1/\nu_{ei} < 1/H < 1/\nu_{e\gamma}$ and only $ 1/\nu_{ei}$ and $1/H$ are relevant time scales. In this case, the photon fails to cool the electrons as energy transfers from DPDM to the electron ion plasma, and the temperature of the electrons can grow significantly. However, the electron ion scattering rate $\nu_{ei}$ decreases as $T_e$ increases, that is, non-resonant heating slows down. This allows us to compute a terminal temperature $T_{\rm final}$ of the electrons, ions and neutral atoms (in thermal equilibrium) as a function of $m_{A'}$ and $\varepsilon$, which we present in Fig.~\ref{fig:terminalT}. During dark ages, which we take to be for $20 < z < 500$ such that the electron photon scattering rate is comparable or smaller than the Hubble rate, the terminal temperature shall be smaller than $\sim 10\,{\rm eV}$ to avoid collisional reionization~\cite{McDermott:2019lch}. This translates to a conservative constraint (orange shaded region in Fig.~\ref{fig:money}) from the dark ages. Such a limit is weaker than the constraints from gas clouds relying on similar non-resonant conversion~\cite{Dubovsky:2015cca,Wadekar:2019mpc}, which remains true even if we extend the redshift range to as early as $z=800$. 
A similar $T_{\rm final}$ can be computed also for the redshift range of $2\lesssim z \lesssim 6$, where similar indirect information about $T_{\rm final}$ can be derived from Helium reionization ($T_{\rm final} \lesssim 40\,{\rm eV}$), and direct information can be derived from Lyman-$\alpha$ line width ($T_{\rm final} \lesssim 0.8 \,{\rm eV}$). The resulting constraints on DPDM, similarly, are comparable or weaker than the gas cloud heating bounds, as was also found in~\cite{McDermott:2019lch,Witte:2020rvb}. Future measurements of the global $y$-distortion signal~\cite{Hill:2015tqa,Chluba:2016bvg}, thermal Sunyeav Zeldovich (tSZ) effect~\cite{Coulton:2019ign}, or 21cm signal~\cite{Sun:2025ksr} might have better sensitivity to this late time energy injection. We leave these studies to future work.

\end{appendix}

\bibliographystyle{apsrev4-2} 
\bibliography{Ref}

\begin{thebibliography}{89}%
\makeatletter
\providecommand \@ifxundefined [1]{%
 \@ifx{#1\undefined}
}%
\providecommand \@ifnum [1]{%
 \ifnum #1\expandafter \@firstoftwo
 \else \expandafter \@secondoftwo
 \fi
}%
\providecommand \@ifx [1]{%
 \ifx #1\expandafter \@firstoftwo
 \else \expandafter \@secondoftwo
 \fi
}%
\providecommand \natexlab [1]{#1}%
\providecommand \enquote  [1]{``#1''}%
\providecommand \bibnamefont  [1]{#1}%
\providecommand \bibfnamefont [1]{#1}%
\providecommand \citenamefont [1]{#1}%
\providecommand \href@noop [0]{\@secondoftwo}%
\providecommand \href [0]{\begingroup \@sanitize@url \@href}%
\providecommand \@href[1]{\@@startlink{#1}\@@href}%
\providecommand \@@href[1]{\endgroup#1\@@endlink}%
\providecommand \@sanitize@url [0]{\catcode `\\12\catcode `\$12\catcode `\&12\catcode `\#12\catcode `\^12\catcode `\_12\catcode `\%12\relax}%
\providecommand \@@startlink[1]{}%
\providecommand \@@endlink[0]{}%
\providecommand \url  [0]{\begingroup\@sanitize@url \@url }%
\providecommand \@url [1]{\endgroup\@href {#1}{\urlprefix }}%
\providecommand \urlprefix  [0]{URL }%
\providecommand \Eprint [0]{\href }%
\providecommand \doibase [0]{https://doi.org/}%
\providecommand \selectlanguage [0]{\@gobble}%
\providecommand \bibinfo  [0]{\@secondoftwo}%
\providecommand \bibfield  [0]{\@secondoftwo}%
\providecommand \translation [1]{[#1]}%
\providecommand \BibitemOpen [0]{}%
\providecommand \bibitemStop [0]{}%
\providecommand \bibitemNoStop [0]{.\EOS\space}%
\providecommand \EOS [0]{\spacefactor3000\relax}%
\providecommand \BibitemShut  [1]{\csname bibitem#1\endcsname}%
\let\auto@bib@innerbib\@empty
\bibitem [{\citenamefont {{Rubin}}\ and\ \citenamefont {{Ford}}(1970)}]{1970ApJ...159..379R}%
  \BibitemOpen
  \bibfield  {author} {\bibinfo {author} {\bibfnamefont {V.~C.}\ \bibnamefont {{Rubin}}}\ and\ \bibinfo {author} {\bibfnamefont {W.~K.}\ \bibnamefont {{Ford}}, \bibfnamefont {Jr.}},\ }\href {https://doi.org/10.1086/150317} {\bibfield  {journal} {\bibinfo  {journal} {\apj}\ }\textbf {\bibinfo {volume} {159}},\ \bibinfo {pages} {379} (\bibinfo {year} {1970})}\BibitemShut {NoStop}%
\bibitem [{\citenamefont {Bergstr{\"o}m}(2000)}]{Bergstrom:2000pn}%
  \BibitemOpen
  \bibfield  {author} {\bibinfo {author} {\bibfnamefont {L.}~\bibnamefont {Bergstr{\"o}m}},\ }\href {https://doi.org/10.1088/0034-4885/63/5/2r3} {\bibfield  {journal} {\bibinfo  {journal} {Rept. Prog. Phys.}\ }\textbf {\bibinfo {volume} {63}},\ \bibinfo {pages} {793} (\bibinfo {year} {2000})},\ \Eprint {https://arxiv.org/abs/hep-ph/0002126} {arXiv:hep-ph/0002126} \BibitemShut {NoStop}%
\bibitem [{\citenamefont {Bertone}\ \emph {et~al.}(2005)\citenamefont {Bertone}, \citenamefont {Hooper},\ and\ \citenamefont {Silk}}]{Bertone:2004pz}%
  \BibitemOpen
  \bibfield  {author} {\bibinfo {author} {\bibfnamefont {G.}~\bibnamefont {Bertone}}, \bibinfo {author} {\bibfnamefont {D.}~\bibnamefont {Hooper}},\ and\ \bibinfo {author} {\bibfnamefont {J.}~\bibnamefont {Silk}},\ }\href {https://doi.org/10.1016/j.physrep.2004.08.031} {\bibfield  {journal} {\bibinfo  {journal} {Phys. Rept.}\ }\textbf {\bibinfo {volume} {405}},\ \bibinfo {pages} {279} (\bibinfo {year} {2005})},\ \Eprint {https://arxiv.org/abs/hep-ph/0404175} {arXiv:hep-ph/0404175} \BibitemShut {NoStop}%
\bibitem [{\citenamefont {Aghanim}\ \emph {et~al.}(2020)\citenamefont {Aghanim} \emph {et~al.}}]{Planck:2018vyg}%
  \BibitemOpen
  \bibfield  {author} {\bibinfo {author} {\bibfnamefont {N.}~\bibnamefont {Aghanim}} \emph {et~al.} (\bibinfo {collaboration} {Planck}),\ }\href {https://doi.org/10.1051/0004-6361/201833910} {\bibfield  {journal} {\bibinfo  {journal} {Astron. Astrophys.}\ }\textbf {\bibinfo {volume} {641}},\ \bibinfo {pages} {A6} (\bibinfo {year} {2020})},\ \bibinfo {note} {[Erratum: Astron.Astrophys. 652, C4 (2021)]},\ \Eprint {https://arxiv.org/abs/1807.06209} {arXiv:1807.06209 [astro-ph.CO]} \BibitemShut {NoStop}%
\bibitem [{\citenamefont {Svrcek}\ and\ \citenamefont {Witten}(2006)}]{Svrcek:2006yi}%
  \BibitemOpen
  \bibfield  {author} {\bibinfo {author} {\bibfnamefont {P.}~\bibnamefont {Svrcek}}\ and\ \bibinfo {author} {\bibfnamefont {E.}~\bibnamefont {Witten}},\ }\href {https://doi.org/10.1088/1126-6708/2006/06/051} {\bibfield  {journal} {\bibinfo  {journal} {JHEP}\ }\textbf {\bibinfo {volume} {06}},\ \bibinfo {pages} {051}},\ \Eprint {https://arxiv.org/abs/hep-th/0605206} {arXiv:hep-th/0605206} \BibitemShut {NoStop}%
\bibitem [{\citenamefont {Arvanitaki}\ \emph {et~al.}(2010)\citenamefont {Arvanitaki}, \citenamefont {Dimopoulos}, \citenamefont {Dubovsky}, \citenamefont {Kaloper},\ and\ \citenamefont {March-Russell}}]{Arvanitaki:2009fg}%
  \BibitemOpen
  \bibfield  {author} {\bibinfo {author} {\bibfnamefont {A.}~\bibnamefont {Arvanitaki}}, \bibinfo {author} {\bibfnamefont {S.}~\bibnamefont {Dimopoulos}}, \bibinfo {author} {\bibfnamefont {S.}~\bibnamefont {Dubovsky}}, \bibinfo {author} {\bibfnamefont {N.}~\bibnamefont {Kaloper}},\ and\ \bibinfo {author} {\bibfnamefont {J.}~\bibnamefont {March-Russell}},\ }\href {https://doi.org/10.1103/PhysRevD.81.123530} {\bibfield  {journal} {\bibinfo  {journal} {Phys. Rev. D}\ }\textbf {\bibinfo {volume} {81}},\ \bibinfo {pages} {123530} (\bibinfo {year} {2010})},\ \Eprint {https://arxiv.org/abs/0905.4720} {arXiv:0905.4720 [hep-th]} \BibitemShut {NoStop}%
\bibitem [{\citenamefont {Goodsell}\ \emph {et~al.}(2009)\citenamefont {Goodsell}, \citenamefont {Jaeckel}, \citenamefont {Redondo},\ and\ \citenamefont {Ringwald}}]{Goodsell:2009xc}%
  \BibitemOpen
  \bibfield  {author} {\bibinfo {author} {\bibfnamefont {M.}~\bibnamefont {Goodsell}}, \bibinfo {author} {\bibfnamefont {J.}~\bibnamefont {Jaeckel}}, \bibinfo {author} {\bibfnamefont {J.}~\bibnamefont {Redondo}},\ and\ \bibinfo {author} {\bibfnamefont {A.}~\bibnamefont {Ringwald}},\ }\href {https://doi.org/10.1088/1126-6708/2009/11/027} {\bibfield  {journal} {\bibinfo  {journal} {JHEP}\ }\textbf {\bibinfo {volume} {11}},\ \bibinfo {pages} {027}},\ \Eprint {https://arxiv.org/abs/0909.0515} {arXiv:0909.0515 [hep-ph]} \BibitemShut {NoStop}%
\bibitem [{\citenamefont {Kaluza}(1921)}]{Kaluza:1921tu}%
  \BibitemOpen
  \bibfield  {author} {\bibinfo {author} {\bibfnamefont {T.}~\bibnamefont {Kaluza}},\ }\href {https://doi.org/10.1142/S0218271818700017} {\bibfield  {journal} {\bibinfo  {journal} {Sitzungsber. Preuss. Akad. Wiss. Berlin (Math. Phys. )}\ }\textbf {\bibinfo {volume} {1921}},\ \bibinfo {pages} {966} (\bibinfo {year} {1921})},\ \Eprint {https://arxiv.org/abs/1803.08616} {arXiv:1803.08616 [physics.hist-ph]} \BibitemShut {NoStop}%
\bibitem [{\citenamefont {Klein}(1926)}]{Klein:1926tv}%
  \BibitemOpen
  \bibfield  {author} {\bibinfo {author} {\bibfnamefont {O.}~\bibnamefont {Klein}},\ }\href {https://doi.org/10.1007/BF01397481} {\bibfield  {journal} {\bibinfo  {journal} {Z. Phys.}\ }\textbf {\bibinfo {volume} {37}},\ \bibinfo {pages} {895} (\bibinfo {year} {1926})}\BibitemShut {NoStop}%
\bibitem [{\citenamefont {Preskill}\ \emph {et~al.}(1983)\citenamefont {Preskill}, \citenamefont {Wise},\ and\ \citenamefont {Wilczek}}]{Preskill:1982cy}%
  \BibitemOpen
  \bibfield  {author} {\bibinfo {author} {\bibfnamefont {J.}~\bibnamefont {Preskill}}, \bibinfo {author} {\bibfnamefont {M.~B.}\ \bibnamefont {Wise}},\ and\ \bibinfo {author} {\bibfnamefont {F.}~\bibnamefont {Wilczek}},\ }\href {https://doi.org/10.1016/0370-2693(83)90637-8} {\bibfield  {journal} {\bibinfo  {journal} {Phys. Lett.}\ }\textbf {\bibinfo {volume} {B120}},\ \bibinfo {pages} {127} (\bibinfo {year} {1983})}\BibitemShut {NoStop}%
\bibitem [{\citenamefont {Redondo}\ and\ \citenamefont {Postma}(2009)}]{Redondo:2008ec}%
  \BibitemOpen
  \bibfield  {author} {\bibinfo {author} {\bibfnamefont {J.}~\bibnamefont {Redondo}}\ and\ \bibinfo {author} {\bibfnamefont {M.}~\bibnamefont {Postma}},\ }\href {https://doi.org/10.1088/1475-7516/2009/02/005} {\bibfield  {journal} {\bibinfo  {journal} {JCAP}\ }\textbf {\bibinfo {volume} {02}},\ \bibinfo {pages} {005}},\ \Eprint {https://arxiv.org/abs/0811.0326} {arXiv:0811.0326 [hep-ph]} \BibitemShut {NoStop}%
\bibitem [{\citenamefont {Nelson}\ and\ \citenamefont {Scholtz}(2011)}]{Nelson:2011sf}%
  \BibitemOpen
  \bibfield  {author} {\bibinfo {author} {\bibfnamefont {A.~E.}\ \bibnamefont {Nelson}}\ and\ \bibinfo {author} {\bibfnamefont {J.}~\bibnamefont {Scholtz}},\ }\href {https://doi.org/10.1103/PhysRevD.84.103501} {\bibfield  {journal} {\bibinfo  {journal} {Phys. Rev. D}\ }\textbf {\bibinfo {volume} {84}},\ \bibinfo {pages} {103501} (\bibinfo {year} {2011})},\ \Eprint {https://arxiv.org/abs/1105.2812} {arXiv:1105.2812 [hep-ph]} \BibitemShut {NoStop}%
\bibitem [{\citenamefont {Graham}\ \emph {et~al.}(2016)\citenamefont {Graham}, \citenamefont {Mardon},\ and\ \citenamefont {Rajendran}}]{Graham:2015rva}%
  \BibitemOpen
  \bibfield  {author} {\bibinfo {author} {\bibfnamefont {P.~W.}\ \bibnamefont {Graham}}, \bibinfo {author} {\bibfnamefont {J.}~\bibnamefont {Mardon}},\ and\ \bibinfo {author} {\bibfnamefont {S.}~\bibnamefont {Rajendran}},\ }\href {https://doi.org/10.1103/PhysRevD.93.103520} {\bibfield  {journal} {\bibinfo  {journal} {Phys. Rev. D}\ }\textbf {\bibinfo {volume} {93}},\ \bibinfo {pages} {103520} (\bibinfo {year} {2016})},\ \Eprint {https://arxiv.org/abs/1504.02102} {arXiv:1504.02102 [hep-ph]} \BibitemShut {NoStop}%
\bibitem [{\citenamefont {Gorghetto}\ \emph {et~al.}(2021)\citenamefont {Gorghetto}, \citenamefont {Hardy},\ and\ \citenamefont {Villadoro}}]{Gorghetto:2020qws}%
  \BibitemOpen
  \bibfield  {author} {\bibinfo {author} {\bibfnamefont {M.}~\bibnamefont {Gorghetto}}, \bibinfo {author} {\bibfnamefont {E.}~\bibnamefont {Hardy}},\ and\ \bibinfo {author} {\bibfnamefont {G.}~\bibnamefont {Villadoro}},\ }\href {https://doi.org/10.21468/SciPostPhys.10.2.050} {\bibfield  {journal} {\bibinfo  {journal} {SciPost Phys.}\ }\textbf {\bibinfo {volume} {10}},\ \bibinfo {pages} {050} (\bibinfo {year} {2021})},\ \Eprint {https://arxiv.org/abs/2007.04990} {arXiv:2007.04990 [hep-ph]} \BibitemShut {NoStop}%
\bibitem [{\citenamefont {Holdom}(1986)}]{Holdom:1985ag}%
  \BibitemOpen
  \bibfield  {author} {\bibinfo {author} {\bibfnamefont {B.}~\bibnamefont {Holdom}},\ }\href {https://doi.org/10.1016/0370-2693(86)91377-8} {\bibfield  {journal} {\bibinfo  {journal} {Phys. Lett.}\ }\textbf {\bibinfo {volume} {B166}},\ \bibinfo {pages} {196} (\bibinfo {year} {1986})}\BibitemShut {NoStop}%
\bibitem [{\citenamefont {Okun}(1982)}]{Okun:1982xi}%
  \BibitemOpen
  \bibfield  {author} {\bibinfo {author} {\bibfnamefont {L.~B.}\ \bibnamefont {Okun}},\ }\href@noop {} {\bibfield  {journal} {\bibinfo  {journal} {Sov. Phys. JETP}\ }\textbf {\bibinfo {volume} {56}},\ \bibinfo {pages} {502} (\bibinfo {year} {1982})}\BibitemShut {NoStop}%
\bibitem [{\citenamefont {Chaudhuri}\ \emph {et~al.}(2015)\citenamefont {Chaudhuri}, \citenamefont {Graham}, \citenamefont {Irwin}, \citenamefont {Mardon}, \citenamefont {Rajendran},\ and\ \citenamefont {Zhao}}]{Chaudhuri:2014dla}%
  \BibitemOpen
  \bibfield  {author} {\bibinfo {author} {\bibfnamefont {S.}~\bibnamefont {Chaudhuri}}, \bibinfo {author} {\bibfnamefont {P.~W.}\ \bibnamefont {Graham}}, \bibinfo {author} {\bibfnamefont {K.}~\bibnamefont {Irwin}}, \bibinfo {author} {\bibfnamefont {J.}~\bibnamefont {Mardon}}, \bibinfo {author} {\bibfnamefont {S.}~\bibnamefont {Rajendran}},\ and\ \bibinfo {author} {\bibfnamefont {Y.}~\bibnamefont {Zhao}},\ }\href {https://doi.org/10.1103/PhysRevD.92.075012} {\bibfield  {journal} {\bibinfo  {journal} {Phys. Rev. D}\ }\textbf {\bibinfo {volume} {92}},\ \bibinfo {pages} {075012} (\bibinfo {year} {2015})},\ \Eprint {https://arxiv.org/abs/1411.7382} {arXiv:1411.7382 [hep-ph]} \BibitemShut {NoStop}%
\bibitem [{\citenamefont {Baryakhtar}\ \emph {et~al.}(2018)\citenamefont {Baryakhtar}, \citenamefont {Huang},\ and\ \citenamefont {Lasenby}}]{Baryakhtar:2018doz}%
  \BibitemOpen
  \bibfield  {author} {\bibinfo {author} {\bibfnamefont {M.}~\bibnamefont {Baryakhtar}}, \bibinfo {author} {\bibfnamefont {J.}~\bibnamefont {Huang}},\ and\ \bibinfo {author} {\bibfnamefont {R.}~\bibnamefont {Lasenby}},\ }\href {https://doi.org/10.1103/PhysRevD.98.035006} {\bibfield  {journal} {\bibinfo  {journal} {Phys. Rev. D}\ }\textbf {\bibinfo {volume} {98}},\ \bibinfo {pages} {035006} (\bibinfo {year} {2018})},\ \Eprint {https://arxiv.org/abs/1803.11455} {arXiv:1803.11455 [hep-ph]} \BibitemShut {NoStop}%
\bibitem [{\citenamefont {Aralis}\ \emph {et~al.}(2020)\citenamefont {Aralis} \emph {et~al.}}]{SuperCDMS:2019jxx}%
  \BibitemOpen
  \bibfield  {author} {\bibinfo {author} {\bibfnamefont {T.}~\bibnamefont {Aralis}} \emph {et~al.} (\bibinfo {collaboration} {SuperCDMS}),\ }\href {https://doi.org/10.1103/PhysRevD.101.052008} {\bibfield  {journal} {\bibinfo  {journal} {Phys. Rev. D}\ }\textbf {\bibinfo {volume} {101}},\ \bibinfo {pages} {052008} (\bibinfo {year} {2020})},\ \bibinfo {note} {[Erratum: Phys.Rev.D 103, 039901 (2021)]},\ \Eprint {https://arxiv.org/abs/1911.11905} {arXiv:1911.11905 [hep-ex]} \BibitemShut {NoStop}%
\bibitem [{\citenamefont {Andrianavalomahefa}\ \emph {et~al.}(2020)\citenamefont {Andrianavalomahefa} \emph {et~al.}}]{FUNKExperiment:2020ofv}%
  \BibitemOpen
  \bibfield  {author} {\bibinfo {author} {\bibfnamefont {A.}~\bibnamefont {Andrianavalomahefa}} \emph {et~al.} (\bibinfo {collaboration} {FUNK Experiment}),\ }\href {https://doi.org/10.1103/PhysRevD.102.042001} {\bibfield  {journal} {\bibinfo  {journal} {Phys. Rev. D}\ }\textbf {\bibinfo {volume} {102}},\ \bibinfo {pages} {042001} (\bibinfo {year} {2020})},\ \Eprint {https://arxiv.org/abs/2003.13144} {arXiv:2003.13144 [astro-ph.CO]} \BibitemShut {NoStop}%
\bibitem [{\citenamefont {Barak}\ \emph {et~al.}(2020)\citenamefont {Barak} \emph {et~al.}}]{SENSEI:2020dpa}%
  \BibitemOpen
  \bibfield  {author} {\bibinfo {author} {\bibfnamefont {L.}~\bibnamefont {Barak}} \emph {et~al.} (\bibinfo {collaboration} {SENSEI}),\ }\href {https://doi.org/10.1103/PhysRevLett.125.171802} {\bibfield  {journal} {\bibinfo  {journal} {Phys. Rev. Lett.}\ }\textbf {\bibinfo {volume} {125}},\ \bibinfo {pages} {171802} (\bibinfo {year} {2020})},\ \Eprint {https://arxiv.org/abs/2004.11378} {arXiv:2004.11378 [astro-ph.CO]} \BibitemShut {NoStop}%
\bibitem [{\citenamefont {Chiles}\ \emph {et~al.}(2022)\citenamefont {Chiles} \emph {et~al.}}]{Chiles:2021gxk}%
  \BibitemOpen
  \bibfield  {author} {\bibinfo {author} {\bibfnamefont {J.}~\bibnamefont {Chiles}} \emph {et~al.},\ }\href {https://doi.org/10.1103/PhysRevLett.128.231802} {\bibfield  {journal} {\bibinfo  {journal} {Phys. Rev. Lett.}\ }\textbf {\bibinfo {volume} {128}},\ \bibinfo {pages} {231802} (\bibinfo {year} {2022})},\ \Eprint {https://arxiv.org/abs/2110.01582} {arXiv:2110.01582 [hep-ex]} \BibitemShut {NoStop}%
\bibitem [{\citenamefont {Cervantes}\ \emph {et~al.}(2022)\citenamefont {Cervantes} \emph {et~al.}}]{Cervantes:2022yzp}%
  \BibitemOpen
  \bibfield  {author} {\bibinfo {author} {\bibfnamefont {R.}~\bibnamefont {Cervantes}} \emph {et~al.},\ }\href {https://doi.org/10.1103/PhysRevLett.129.201301} {\bibfield  {journal} {\bibinfo  {journal} {Phys. Rev. Lett.}\ }\textbf {\bibinfo {volume} {129}},\ \bibinfo {pages} {201301} (\bibinfo {year} {2022})},\ \Eprint {https://arxiv.org/abs/2204.03818} {arXiv:2204.03818 [hep-ex]} \BibitemShut {NoStop}%
\bibitem [{\citenamefont {Fixsen}\ \emph {et~al.}(1996)\citenamefont {Fixsen}, \citenamefont {Cheng}, \citenamefont {Gales}, \citenamefont {Mather}, \citenamefont {Shafer},\ and\ \citenamefont {Wright}}]{Fixsen:1996nj}%
  \BibitemOpen
  \bibfield  {author} {\bibinfo {author} {\bibfnamefont {D.~J.}\ \bibnamefont {Fixsen}}, \bibinfo {author} {\bibfnamefont {E.~S.}\ \bibnamefont {Cheng}}, \bibinfo {author} {\bibfnamefont {J.~M.}\ \bibnamefont {Gales}}, \bibinfo {author} {\bibfnamefont {J.~C.}\ \bibnamefont {Mather}}, \bibinfo {author} {\bibfnamefont {R.~A.}\ \bibnamefont {Shafer}},\ and\ \bibinfo {author} {\bibfnamefont {E.~L.}\ \bibnamefont {Wright}},\ }\href {https://doi.org/10.1086/178173} {\bibfield  {journal} {\bibinfo  {journal} {Astrophys. J.}\ }\textbf {\bibinfo {volume} {473}},\ \bibinfo {pages} {576} (\bibinfo {year} {1996})},\ \Eprint {https://arxiv.org/abs/astro-ph/9605054} {arXiv:astro-ph/9605054} \BibitemShut {NoStop}%
\bibitem [{\citenamefont {Mirizzi}\ \emph {et~al.}(2009)\citenamefont {Mirizzi}, \citenamefont {Redondo},\ and\ \citenamefont {Sigl}}]{Mirizzi:2009iz}%
  \BibitemOpen
  \bibfield  {author} {\bibinfo {author} {\bibfnamefont {A.}~\bibnamefont {Mirizzi}}, \bibinfo {author} {\bibfnamefont {J.}~\bibnamefont {Redondo}},\ and\ \bibinfo {author} {\bibfnamefont {G.}~\bibnamefont {Sigl}},\ }\href {https://doi.org/10.1088/1475-7516/2009/03/026} {\bibfield  {journal} {\bibinfo  {journal} {JCAP}\ }\textbf {\bibinfo {volume} {03}},\ \bibinfo {pages} {026}},\ \Eprint {https://arxiv.org/abs/0901.0014} {arXiv:0901.0014 [hep-ph]} \BibitemShut {NoStop}%
\bibitem [{\citenamefont {Kunze}\ and\ \citenamefont {V{\'a}zquez-Mozo}(2015)}]{Kunze:2015noa}%
  \BibitemOpen
  \bibfield  {author} {\bibinfo {author} {\bibfnamefont {K.~E.}\ \bibnamefont {Kunze}}\ and\ \bibinfo {author} {\bibfnamefont {M.~{\'A}.}\ \bibnamefont {V{\'a}zquez-Mozo}},\ }\href {https://doi.org/10.1088/1475-7516/2015/12/028} {\bibfield  {journal} {\bibinfo  {journal} {JCAP}\ }\textbf {\bibinfo {volume} {12}},\ \bibinfo {pages} {028}},\ \Eprint {https://arxiv.org/abs/1507.02614} {arXiv:1507.02614 [astro-ph.CO]} \BibitemShut {NoStop}%
\bibitem [{\citenamefont {McDermott}\ and\ \citenamefont {Witte}(2020)}]{McDermott:2019lch}%
  \BibitemOpen
  \bibfield  {author} {\bibinfo {author} {\bibfnamefont {S.~D.}\ \bibnamefont {McDermott}}\ and\ \bibinfo {author} {\bibfnamefont {S.~J.}\ \bibnamefont {Witte}},\ }\href {https://doi.org/10.1103/PhysRevD.101.063030} {\bibfield  {journal} {\bibinfo  {journal} {Phys. Rev. D}\ }\textbf {\bibinfo {volume} {101}},\ \bibinfo {pages} {063030} (\bibinfo {year} {2020})},\ \Eprint {https://arxiv.org/abs/1911.05086} {arXiv:1911.05086 [hep-ph]} \BibitemShut {NoStop}%
\bibitem [{\citenamefont {Caputo}\ \emph {et~al.}(2020{\natexlab{a}})\citenamefont {Caputo}, \citenamefont {Liu}, \citenamefont {Mishra-Sharma},\ and\ \citenamefont {Ruderman}}]{Caputo:2020rnx}%
  \BibitemOpen
  \bibfield  {author} {\bibinfo {author} {\bibfnamefont {A.}~\bibnamefont {Caputo}}, \bibinfo {author} {\bibfnamefont {H.}~\bibnamefont {Liu}}, \bibinfo {author} {\bibfnamefont {S.}~\bibnamefont {Mishra-Sharma}},\ and\ \bibinfo {author} {\bibfnamefont {J.~T.}\ \bibnamefont {Ruderman}},\ }\href {https://doi.org/10.1103/PhysRevD.102.103533} {\bibfield  {journal} {\bibinfo  {journal} {Phys. Rev. D}\ }\textbf {\bibinfo {volume} {102}},\ \bibinfo {pages} {103533} (\bibinfo {year} {2020}{\natexlab{a}})},\ \Eprint {https://arxiv.org/abs/2004.06733} {arXiv:2004.06733 [astro-ph.CO]} \BibitemShut {NoStop}%
\bibitem [{\citenamefont {Caputo}\ \emph {et~al.}(2020{\natexlab{b}})\citenamefont {Caputo}, \citenamefont {Liu}, \citenamefont {Mishra-Sharma},\ and\ \citenamefont {Ruderman}}]{Caputo:2020bdy}%
  \BibitemOpen
  \bibfield  {author} {\bibinfo {author} {\bibfnamefont {A.}~\bibnamefont {Caputo}}, \bibinfo {author} {\bibfnamefont {H.}~\bibnamefont {Liu}}, \bibinfo {author} {\bibfnamefont {S.}~\bibnamefont {Mishra-Sharma}},\ and\ \bibinfo {author} {\bibfnamefont {J.~T.}\ \bibnamefont {Ruderman}},\ }\href {https://doi.org/10.1103/PhysRevLett.125.221303} {\bibfield  {journal} {\bibinfo  {journal} {Phys. Rev. Lett.}\ }\textbf {\bibinfo {volume} {125}},\ \bibinfo {pages} {221303} (\bibinfo {year} {2020}{\natexlab{b}})},\ \Eprint {https://arxiv.org/abs/2002.05165} {arXiv:2002.05165 [astro-ph.CO]} \BibitemShut {NoStop}%
\bibitem [{\citenamefont {P{\textasciicircum}{\i}rvu}\ \emph {et~al.}(2024)\citenamefont {P{\textasciicircum}{\i}rvu}, \citenamefont {Huang},\ and\ \citenamefont {Johnson}}]{Pirvu:2023lch}%
  \BibitemOpen
  \bibfield  {author} {\bibinfo {author} {\bibfnamefont {D.}~\bibnamefont {P{\textasciicircum}{\i}rvu}}, \bibinfo {author} {\bibfnamefont {J.}~\bibnamefont {Huang}},\ and\ \bibinfo {author} {\bibfnamefont {M.~C.}\ \bibnamefont {Johnson}},\ }\href {https://doi.org/10.1088/1475-7516/2024/01/019} {\bibfield  {journal} {\bibinfo  {journal} {JCAP}\ }\textbf {\bibinfo {volume} {01}},\ \bibinfo {pages} {019}},\ \Eprint {https://arxiv.org/abs/2307.15124} {arXiv:2307.15124 [hep-ph]} \BibitemShut {NoStop}%
\bibitem [{\citenamefont {Siemonsen}\ \emph {et~al.}(2023)\citenamefont {Siemonsen}, \citenamefont {Mondino}, \citenamefont {Egana-Ugrinovic}, \citenamefont {Huang}, \citenamefont {Baryakhtar},\ and\ \citenamefont {East}}]{Siemonsen:2022ivj}%
  \BibitemOpen
  \bibfield  {author} {\bibinfo {author} {\bibfnamefont {N.}~\bibnamefont {Siemonsen}}, \bibinfo {author} {\bibfnamefont {C.}~\bibnamefont {Mondino}}, \bibinfo {author} {\bibfnamefont {D.}~\bibnamefont {Egana-Ugrinovic}}, \bibinfo {author} {\bibfnamefont {J.}~\bibnamefont {Huang}}, \bibinfo {author} {\bibfnamefont {M.}~\bibnamefont {Baryakhtar}},\ and\ \bibinfo {author} {\bibfnamefont {W.~E.}\ \bibnamefont {East}},\ }\href {https://doi.org/10.1103/PhysRevD.107.075025} {\bibfield  {journal} {\bibinfo  {journal} {Phys. Rev. D}\ }\textbf {\bibinfo {volume} {107}},\ \bibinfo {pages} {075025} (\bibinfo {year} {2023})},\ \Eprint {https://arxiv.org/abs/2212.09772} {arXiv:2212.09772 [astro-ph.HE]} \BibitemShut {NoStop}%
\bibitem [{\citenamefont {McCarthy}\ \emph {et~al.}(2024)\citenamefont {McCarthy}, \citenamefont {Pirvu}, \citenamefont {Hill}, \citenamefont {Huang}, \citenamefont {Johnson},\ and\ \citenamefont {Rogers}}]{McCarthy:2024ozh}%
  \BibitemOpen
  \bibfield  {author} {\bibinfo {author} {\bibfnamefont {F.}~\bibnamefont {McCarthy}}, \bibinfo {author} {\bibfnamefont {D.}~\bibnamefont {Pirvu}}, \bibinfo {author} {\bibfnamefont {J.~C.}\ \bibnamefont {Hill}}, \bibinfo {author} {\bibfnamefont {J.}~\bibnamefont {Huang}}, \bibinfo {author} {\bibfnamefont {M.~C.}\ \bibnamefont {Johnson}},\ and\ \bibinfo {author} {\bibfnamefont {K.~K.}\ \bibnamefont {Rogers}},\ }\href {https://doi.org/10.1103/PhysRevLett.133.141003} {\bibfield  {journal} {\bibinfo  {journal} {Phys. Rev. Lett.}\ }\textbf {\bibinfo {volume} {133}},\ \bibinfo {pages} {141003} (\bibinfo {year} {2024})},\ \Eprint {https://arxiv.org/abs/2406.02546} {arXiv:2406.02546 [hep-ph]} \BibitemShut {NoStop}%
\bibitem [{\citenamefont {Chluba}\ \emph {et~al.}(2024)\citenamefont {Chluba}, \citenamefont {Cyr},\ and\ \citenamefont {Johnson}}]{Chluba:2024wui}%
  \BibitemOpen
  \bibfield  {author} {\bibinfo {author} {\bibfnamefont {J.}~\bibnamefont {Chluba}}, \bibinfo {author} {\bibfnamefont {B.}~\bibnamefont {Cyr}},\ and\ \bibinfo {author} {\bibfnamefont {M.~C.}\ \bibnamefont {Johnson}},\ }\href {https://doi.org/10.1093/mnras/stae2464} {\bibfield  {journal} {\bibinfo  {journal} {Mon. Not. Roy. Astron. Soc.}\ }\textbf {\bibinfo {volume} {535}},\ \bibinfo {pages} {1874} (\bibinfo {year} {2024})},\ \Eprint {https://arxiv.org/abs/2409.12115} {arXiv:2409.12115 [astro-ph.CO]} \BibitemShut {NoStop}%
\bibitem [{\citenamefont {Arias}\ \emph {et~al.}(2012)\citenamefont {Arias}, \citenamefont {Cadamuro}, \citenamefont {Goodsell}, \citenamefont {Jaeckel}, \citenamefont {Redondo},\ and\ \citenamefont {Ringwald}}]{Arias:2012az}%
  \BibitemOpen
  \bibfield  {author} {\bibinfo {author} {\bibfnamefont {P.}~\bibnamefont {Arias}}, \bibinfo {author} {\bibfnamefont {D.}~\bibnamefont {Cadamuro}}, \bibinfo {author} {\bibfnamefont {M.}~\bibnamefont {Goodsell}}, \bibinfo {author} {\bibfnamefont {J.}~\bibnamefont {Jaeckel}}, \bibinfo {author} {\bibfnamefont {J.}~\bibnamefont {Redondo}},\ and\ \bibinfo {author} {\bibfnamefont {A.}~\bibnamefont {Ringwald}},\ }\href {https://doi.org/10.1088/1475-7516/2012/06/013} {\bibfield  {journal} {\bibinfo  {journal} {JCAP}\ }\textbf {\bibinfo {volume} {1206}},\ \bibinfo {pages} {013}},\ \Eprint {https://arxiv.org/abs/1201.5902} {arXiv:1201.5902 [hep-ph]} \BibitemShut {NoStop}%
\bibitem [{\citenamefont {Caputo}\ \emph {et~al.}(2021)\citenamefont {Caputo}, \citenamefont {Millar}, \citenamefont {O'Hare},\ and\ \citenamefont {Vitagliano}}]{Caputo:2021eaa}%
  \BibitemOpen
  \bibfield  {author} {\bibinfo {author} {\bibfnamefont {A.}~\bibnamefont {Caputo}}, \bibinfo {author} {\bibfnamefont {A.~J.}\ \bibnamefont {Millar}}, \bibinfo {author} {\bibfnamefont {C.~A.~J.}\ \bibnamefont {O'Hare}},\ and\ \bibinfo {author} {\bibfnamefont {E.}~\bibnamefont {Vitagliano}},\ }\href {https://doi.org/10.1103/PhysRevD.104.095029} {\bibfield  {journal} {\bibinfo  {journal} {Phys. Rev. D}\ }\textbf {\bibinfo {volume} {104}},\ \bibinfo {pages} {095029} (\bibinfo {year} {2021})},\ \Eprint {https://arxiv.org/abs/2105.04565} {arXiv:2105.04565 [hep-ph]} \BibitemShut {NoStop}%
\bibitem [{\citenamefont {Stueckelberg}(1938)}]{Stueckelberg:1938hvi}%
  \BibitemOpen
  \bibfield  {author} {\bibinfo {author} {\bibfnamefont {E.~C.~G.}\ \bibnamefont {Stueckelberg}},\ }\href {https://doi.org/10.5169/seals-110852} {\bibfield  {journal} {\bibinfo  {journal} {Helv. Phys. Acta}\ }\textbf {\bibinfo {volume} {11}},\ \bibinfo {pages} {225} (\bibinfo {year} {1938})}\BibitemShut {NoStop}%
\bibitem [{\citenamefont {Ruegg}\ and\ \citenamefont {Ruiz-Altaba}(2004)}]{Ruegg:2003ps}%
  \BibitemOpen
  \bibfield  {author} {\bibinfo {author} {\bibfnamefont {H.}~\bibnamefont {Ruegg}}\ and\ \bibinfo {author} {\bibfnamefont {M.}~\bibnamefont {Ruiz-Altaba}},\ }\href {https://doi.org/10.1142/S0217751X04019755} {\bibfield  {journal} {\bibinfo  {journal} {Int. J. Mod. Phys. A}\ }\textbf {\bibinfo {volume} {19}},\ \bibinfo {pages} {3265} (\bibinfo {year} {2004})},\ \Eprint {https://arxiv.org/abs/hep-th/0304245} {arXiv:hep-th/0304245} \BibitemShut {NoStop}%
\bibitem [{\citenamefont {East}\ and\ \citenamefont {Huang}(2022)}]{East:2022rsi}%
  \BibitemOpen
  \bibfield  {author} {\bibinfo {author} {\bibfnamefont {W.~E.}\ \bibnamefont {East}}\ and\ \bibinfo {author} {\bibfnamefont {J.}~\bibnamefont {Huang}},\ }\href {https://doi.org/10.1007/JHEP12(2022)089} {\bibfield  {journal} {\bibinfo  {journal} {JHEP}\ }\textbf {\bibinfo {volume} {2022}}\bibfield  {number} {\bibinfo  {number} { (12)},\ \bibinfo {pages} {089}},\ }\Eprint {https://arxiv.org/abs/2206.12432} {arXiv:2206.12432 [hep-ph]} \BibitemShut {NoStop}%
\bibitem [{\citenamefont {Cyncynates}\ and\ \citenamefont {Weiner}(2025{\natexlab{a}})}]{Cyncynates:2023zwj}%
  \BibitemOpen
  \bibfield  {author} {\bibinfo {author} {\bibfnamefont {D.}~\bibnamefont {Cyncynates}}\ and\ \bibinfo {author} {\bibfnamefont {Z.~J.}\ \bibnamefont {Weiner}},\ }\href {https://doi.org/10.1103/PhysRevLett.134.211002} {\bibfield  {journal} {\bibinfo  {journal} {Phys. Rev. Lett.}\ }\textbf {\bibinfo {volume} {134}},\ \bibinfo {pages} {211002} (\bibinfo {year} {2025}{\natexlab{a}})},\ \Eprint {https://arxiv.org/abs/2310.18397} {arXiv:2310.18397 [hep-ph]} \BibitemShut {NoStop}%
\bibitem [{\citenamefont {Cyncynates}\ and\ \citenamefont {Weiner}(2025{\natexlab{b}})}]{Cyncynates:2024yxm}%
  \BibitemOpen
  \bibfield  {author} {\bibinfo {author} {\bibfnamefont {D.}~\bibnamefont {Cyncynates}}\ and\ \bibinfo {author} {\bibfnamefont {Z.~J.}\ \bibnamefont {Weiner}},\ }\href {https://doi.org/10.1103/PhysRevD.111.103535} {\bibfield  {journal} {\bibinfo  {journal} {Phys. Rev. D}\ }\textbf {\bibinfo {volume} {111}},\ \bibinfo {pages} {103535} (\bibinfo {year} {2025}{\natexlab{b}})},\ \Eprint {https://arxiv.org/abs/2410.14774} {arXiv:2410.14774 [hep-ph]} \BibitemShut {NoStop}%
\bibitem [{\citenamefont {Witte}\ \emph {et~al.}(2020)\citenamefont {Witte}, \citenamefont {Rosauro-Alcaraz}, \citenamefont {McDermott},\ and\ \citenamefont {Poulin}}]{Witte:2020rvb}%
  \BibitemOpen
  \bibfield  {author} {\bibinfo {author} {\bibfnamefont {S.~J.}\ \bibnamefont {Witte}}, \bibinfo {author} {\bibfnamefont {S.}~\bibnamefont {Rosauro-Alcaraz}}, \bibinfo {author} {\bibfnamefont {S.~D.}\ \bibnamefont {McDermott}},\ and\ \bibinfo {author} {\bibfnamefont {V.}~\bibnamefont {Poulin}},\ }\href {https://doi.org/10.1007/JHEP06(2020)132} {\bibfield  {journal} {\bibinfo  {journal} {JHEP}\ }\textbf {\bibinfo {volume} {06}},\ \bibinfo {pages} {132}},\ \Eprint {https://arxiv.org/abs/2003.13698} {arXiv:2003.13698 [astro-ph.CO]} \BibitemShut {NoStop}%
\bibitem [{\citenamefont {Chluba}(2016)}]{Chluba:2016bvg}%
  \BibitemOpen
  \bibfield  {author} {\bibinfo {author} {\bibfnamefont {J.}~\bibnamefont {Chluba}},\ }\href {https://doi.org/10.1093/mnras/stw945} {\bibfield  {journal} {\bibinfo  {journal} {Mon. Not. Roy. Astron. Soc.}\ }\textbf {\bibinfo {volume} {460}},\ \bibinfo {pages} {227} (\bibinfo {year} {2016})},\ \Eprint {https://arxiv.org/abs/1603.02496} {arXiv:1603.02496 [astro-ph.CO]} \BibitemShut {NoStop}%
\bibitem [{\citenamefont {Amin}\ \emph {et~al.}(2022)\citenamefont {Amin}, \citenamefont {Jain}, \citenamefont {Karur},\ and\ \citenamefont {Mocz}}]{Amin:2022pzv}%
  \BibitemOpen
  \bibfield  {author} {\bibinfo {author} {\bibfnamefont {M.~A.}\ \bibnamefont {Amin}}, \bibinfo {author} {\bibfnamefont {M.}~\bibnamefont {Jain}}, \bibinfo {author} {\bibfnamefont {R.}~\bibnamefont {Karur}},\ and\ \bibinfo {author} {\bibfnamefont {P.}~\bibnamefont {Mocz}},\ }\href {https://doi.org/10.1088/1475-7516/2022/08/014} {\bibfield  {journal} {\bibinfo  {journal} {JCAP}\ }\textbf {\bibinfo {volume} {08}}\bibfield  {number} {\bibinfo  {number} { (08)},\ \bibinfo {pages} {014}},\ }\Eprint {https://arxiv.org/abs/2203.11935} {arXiv:2203.11935 [astro-ph.CO]} \BibitemShut {NoStop}%
\bibitem [{\citenamefont {{Zakharov}}\ \emph {et~al.}(1985)\citenamefont {{Zakharov}}, \citenamefont {{Musher}},\ and\ \citenamefont {{Rubenchik}}}]{1985PhR...129..285Z}%
  \BibitemOpen
  \bibfield  {author} {\bibinfo {author} {\bibfnamefont {V.~E.}\ \bibnamefont {{Zakharov}}}, \bibinfo {author} {\bibfnamefont {S.~L.}\ \bibnamefont {{Musher}}},\ and\ \bibinfo {author} {\bibfnamefont {A.~M.}\ \bibnamefont {{Rubenchik}}},\ }\href {https://doi.org/10.1016/0370-1573(85)90040-7} {\bibfield  {journal} {\bibinfo  {journal} {\physrep}\ }\textbf {\bibinfo {volume} {129}},\ \bibinfo {pages} {285} (\bibinfo {year} {1985})}\BibitemShut {NoStop}%
\bibitem [{\citenamefont {Schekochihin}(2017)}]{Schekochihin2017Houches}%
  \BibitemOpen
  \bibfield  {author} {\bibinfo {author} {\bibfnamefont {A.~A.}\ \bibnamefont {Schekochihin}},\ }\href@noop {} {\bibinfo {title} {Lecture notes on kinetic theory and magnetohydrodynamics of plasmas}},\ \bibinfo {howpublished} {Lecture notes, Houches Summer School / Workshop, {IPAG} / {OSUG}} (\bibinfo {year} {2017}),\ \bibinfo {note} {available online: \texttt{https://ipag.osug.fr/\~{}lesurg/plasmas2017/talks/Schekochihin-Houches-2017.pdf}}\BibitemShut {NoStop}%
\bibitem [{\citenamefont {{Kentwell}}\ and\ \citenamefont {{Jones}}(1987)}]{1987PhR...145..319K}%
  \BibitemOpen
  \bibfield  {author} {\bibinfo {author} {\bibfnamefont {G.~W.}\ \bibnamefont {{Kentwell}}}\ and\ \bibinfo {author} {\bibfnamefont {D.~A.}\ \bibnamefont {{Jones}}},\ }\href {https://doi.org/10.1016/0370-1573(87)90063-9} {\bibfield  {journal} {\bibinfo  {journal} {\physrep}\ }\textbf {\bibinfo {volume} {145}},\ \bibinfo {pages} {319} (\bibinfo {year} {1987})}\BibitemShut {NoStop}%
\bibitem [{\citenamefont {{Robinson}}(1997)}]{1997RvMP...69..507R}%
  \BibitemOpen
  \bibfield  {author} {\bibinfo {author} {\bibfnamefont {P.~A.}\ \bibnamefont {{Robinson}}},\ }\href {https://doi.org/10.1103/RevModPhys.69.507} {\bibfield  {journal} {\bibinfo  {journal} {Reviews of Modern Physics}\ }\textbf {\bibinfo {volume} {69}},\ \bibinfo {pages} {507} (\bibinfo {year} {1997})}\BibitemShut {NoStop}%
\bibitem [{\citenamefont {{Kirkwood}}\ \emph {et~al.}(1999)\citenamefont {{Kirkwood}}, \citenamefont {{Montgomery}}, \citenamefont {{Afeyan}}, \citenamefont {{Moody}}, \citenamefont {{MacGowan}}, \citenamefont {{Joshi}}, \citenamefont {{Wharton}}, \citenamefont {{Glenzer}}, \citenamefont {{Williams}}, \citenamefont {{Young}}, \citenamefont {{Kruer}}, \citenamefont {{Estabrook}},\ and\ \citenamefont {{Berger}}}]{1999PhRvL..83.2965K}%
  \BibitemOpen
  \bibfield  {author} {\bibinfo {author} {\bibfnamefont {R.~K.}\ \bibnamefont {{Kirkwood}}}, \bibinfo {author} {\bibfnamefont {D.~S.}\ \bibnamefont {{Montgomery}}}, \bibinfo {author} {\bibfnamefont {B.~B.}\ \bibnamefont {{Afeyan}}}, \bibinfo {author} {\bibfnamefont {J.~D.}\ \bibnamefont {{Moody}}}, \bibinfo {author} {\bibfnamefont {B.~J.}\ \bibnamefont {{MacGowan}}}, \bibinfo {author} {\bibfnamefont {C.}~\bibnamefont {{Joshi}}}, \bibinfo {author} {\bibfnamefont {K.~B.}\ \bibnamefont {{Wharton}}}, \bibinfo {author} {\bibfnamefont {S.~H.}\ \bibnamefont {{Glenzer}}}, \bibinfo {author} {\bibfnamefont {E.~A.}\ \bibnamefont {{Williams}}}, \bibinfo {author} {\bibfnamefont {P.~E.}\ \bibnamefont {{Young}}}, \bibinfo {author} {\bibfnamefont {W.~L.}\ \bibnamefont {{Kruer}}}, \bibinfo {author} {\bibfnamefont {K.~G.}\ \bibnamefont {{Estabrook}}},\ and\ \bibinfo {author} {\bibfnamefont {R.~L.}\ \bibnamefont {{Berger}}},\ }\href {https://doi.org/10.1103/PhysRevLett.83.2965} {\bibfield  {journal} {\bibinfo  {journal} {\prl}\
  }\textbf {\bibinfo {volume} {83}},\ \bibinfo {pages} {2965} (\bibinfo {year} {1999})}\BibitemShut {NoStop}%
\bibitem [{\citenamefont {{Dubois}}\ \emph {et~al.}(1991)\citenamefont {{Dubois}}, \citenamefont {{Rose}},\ and\ \citenamefont {{Russell}}}]{1991PhRvL..66.1970D}%
  \BibitemOpen
  \bibfield  {author} {\bibinfo {author} {\bibfnamefont {D.~F.}\ \bibnamefont {{Dubois}}}, \bibinfo {author} {\bibfnamefont {H.~A.}\ \bibnamefont {{Rose}}},\ and\ \bibinfo {author} {\bibfnamefont {D.}~\bibnamefont {{Russell}}},\ }\href {https://doi.org/10.1103/PhysRevLett.66.1970} {\bibfield  {journal} {\bibinfo  {journal} {\prl}\ }\textbf {\bibinfo {volume} {66}},\ \bibinfo {pages} {1970} (\bibinfo {year} {1991})}\BibitemShut {NoStop}%
\bibitem [{\citenamefont {Russell}\ \emph {et~al.}(1986)\citenamefont {Russell}, \citenamefont {DuBois},\ and\ \citenamefont {Rose}}]{PhysRevLett.56.838}%
  \BibitemOpen
  \bibfield  {author} {\bibinfo {author} {\bibfnamefont {D.}~\bibnamefont {Russell}}, \bibinfo {author} {\bibfnamefont {D.~F.}\ \bibnamefont {DuBois}},\ and\ \bibinfo {author} {\bibfnamefont {H.~A.}\ \bibnamefont {Rose}},\ }\href {https://doi.org/10.1103/PhysRevLett.56.838} {\bibfield  {journal} {\bibinfo  {journal} {Phys. Rev. Lett.}\ }\textbf {\bibinfo {volume} {56}},\ \bibinfo {pages} {838} (\bibinfo {year} {1986})}\BibitemShut {NoStop}%
\bibitem [{\citenamefont {{Russell}}\ \emph {et~al.}(1988)\citenamefont {{Russell}}, \citenamefont {{Dubois}},\ and\ \citenamefont {{Rose}}}]{1988PhRvL..60..581R}%
  \BibitemOpen
  \bibfield  {author} {\bibinfo {author} {\bibfnamefont {D.}~\bibnamefont {{Russell}}}, \bibinfo {author} {\bibfnamefont {D.~F.}\ \bibnamefont {{Dubois}}},\ and\ \bibinfo {author} {\bibfnamefont {H.~A.}\ \bibnamefont {{Rose}}},\ }\href {https://doi.org/10.1103/PhysRevLett.60.581} {\bibfield  {journal} {\bibinfo  {journal} {\prl}\ }\textbf {\bibinfo {volume} {60}},\ \bibinfo {pages} {581} (\bibinfo {year} {1988})}\BibitemShut {NoStop}%
\bibitem [{\citenamefont {Ivanov}\ and\ \citenamefont {Rudakov}(1967)}]{ivanov1967quasilinear}%
  \BibitemOpen
  \bibfield  {author} {\bibinfo {author} {\bibfnamefont {A.}~\bibnamefont {Ivanov}}\ and\ \bibinfo {author} {\bibfnamefont {L.}~\bibnamefont {Rudakov}},\ }\href@noop {} {\bibfield  {journal} {\bibinfo  {journal} {Sov. Phys.—JETP}\ }\textbf {\bibinfo {volume} {24}},\ \bibinfo {pages} {1027} (\bibinfo {year} {1967})}\BibitemShut {NoStop}%
\bibitem [{\citenamefont {{Zakharov}}(1972)}]{1972JETP...35..908Z}%
  \BibitemOpen
  \bibfield  {author} {\bibinfo {author} {\bibfnamefont {V.~E.}\ \bibnamefont {{Zakharov}}},\ }\href@noop {} {\bibfield  {journal} {\bibinfo  {journal} {Soviet Journal of Experimental and Theoretical Physics}\ }\textbf {\bibinfo {volume} {35}},\ \bibinfo {pages} {908} (\bibinfo {year} {1972})}\BibitemShut {NoStop}%
\bibitem [{\citenamefont {{Fried}}\ and\ \citenamefont {{Gould}}(1961)}]{1961PhFl....4..139F}%
  \BibitemOpen
  \bibfield  {author} {\bibinfo {author} {\bibfnamefont {B.~D.}\ \bibnamefont {{Fried}}}\ and\ \bibinfo {author} {\bibfnamefont {R.~W.}\ \bibnamefont {{Gould}}},\ }\href {https://doi.org/10.1063/1.1706174} {\bibfield  {journal} {\bibinfo  {journal} {Physics of Fluids}\ }\textbf {\bibinfo {volume} {4}},\ \bibinfo {pages} {139} (\bibinfo {year} {1961})}\BibitemShut {NoStop}%
\bibitem [{\citenamefont {{Shalaby}}\ \emph {et~al.}(2017{\natexlab{a}})\citenamefont {{Shalaby}}, \citenamefont {{Broderick}}, \citenamefont {{Chang}}, \citenamefont {{Pfrommer}}, \citenamefont {{Lamberts}},\ and\ \citenamefont {{Puchwein}}}]{sharp}%
  \BibitemOpen
  \bibfield  {author} {\bibinfo {author} {\bibfnamefont {M.}~\bibnamefont {{Shalaby}}}, \bibinfo {author} {\bibfnamefont {A.~E.}\ \bibnamefont {{Broderick}}}, \bibinfo {author} {\bibfnamefont {P.}~\bibnamefont {{Chang}}}, \bibinfo {author} {\bibfnamefont {C.}~\bibnamefont {{Pfrommer}}}, \bibinfo {author} {\bibfnamefont {A.}~\bibnamefont {{Lamberts}}},\ and\ \bibinfo {author} {\bibfnamefont {E.}~\bibnamefont {{Puchwein}}},\ }\href {https://doi.org/10.3847/1538-4357/aa6d13} {\bibfield  {journal} {\bibinfo  {journal} {\apj}\ }\textbf {\bibinfo {volume} {841}},\ \bibinfo {eid} {52} (\bibinfo {year} {2017}{\natexlab{a}})},\ \Eprint {https://arxiv.org/abs/1702.04732} {arXiv:1702.04732 [physics.comp-ph]} \BibitemShut {NoStop}%
\bibitem [{\citenamefont {{Shalaby}}\ \emph {et~al.}(2021)\citenamefont {{Shalaby}}, \citenamefont {{Thomas}},\ and\ \citenamefont {{Pfrommer}}}]{sharp2}%
  \BibitemOpen
  \bibfield  {author} {\bibinfo {author} {\bibfnamefont {M.}~\bibnamefont {{Shalaby}}}, \bibinfo {author} {\bibfnamefont {T.}~\bibnamefont {{Thomas}}},\ and\ \bibinfo {author} {\bibfnamefont {C.}~\bibnamefont {{Pfrommer}}},\ }\href {https://doi.org/10.3847/1538-4357/abd02d} {\bibfield  {journal} {\bibinfo  {journal} {\apj}\ }\textbf {\bibinfo {volume} {908}},\ \bibinfo {eid} {206} (\bibinfo {year} {2021})},\ \Eprint {https://arxiv.org/abs/2010.11197} {arXiv:2010.11197 [astro-ph.HE]} \BibitemShut {NoStop}%
\bibitem [{\citenamefont {{Bret}}\ \emph {et~al.}(2010)\citenamefont {{Bret}}, \citenamefont {{Gremillet}},\ and\ \citenamefont {{Dieckmann}}}]{bret2010}%
  \BibitemOpen
  \bibfield  {author} {\bibinfo {author} {\bibfnamefont {A.}~\bibnamefont {{Bret}}}, \bibinfo {author} {\bibfnamefont {L.}~\bibnamefont {{Gremillet}}},\ and\ \bibinfo {author} {\bibfnamefont {M.~E.}\ \bibnamefont {{Dieckmann}}},\ }\href {https://doi.org/10.1063/1.3514586} {\bibfield  {journal} {\bibinfo  {journal} {Physics of Plasmas}\ }\textbf {\bibinfo {volume} {17}},\ \bibinfo {eid} {120501} (\bibinfo {year} {2010})}\BibitemShut {NoStop}%
\bibitem [{\citenamefont {{Blas}}\ \emph {et~al.}(2011)\citenamefont {{Blas}}, \citenamefont {{Lesgourgues}},\ and\ \citenamefont {{Tram}}}]{2011JCAP...07..034B}%
  \BibitemOpen
  \bibfield  {author} {\bibinfo {author} {\bibfnamefont {D.}~\bibnamefont {{Blas}}}, \bibinfo {author} {\bibfnamefont {J.}~\bibnamefont {{Lesgourgues}}},\ and\ \bibinfo {author} {\bibfnamefont {T.}~\bibnamefont {{Tram}}},\ }\href {https://doi.org/10.1088/1475-7516/2011/07/034} {\bibfield  {journal} {\bibinfo  {journal} {\jcap}\ }\textbf {\bibinfo {volume} {2011}},\ \bibinfo {eid} {034} (\bibinfo {year} {2011})},\ \Eprint {https://arxiv.org/abs/1104.2933} {arXiv:1104.2933 [astro-ph.CO]} \BibitemShut {NoStop}%
\bibitem [{Vid(2025)}]{Video_DPDM}%
  \BibitemOpen
  \href {https://mohamadshalaby.github.io/DPDM.html} {\bibinfo {title} {mohamadshalaby.github.io/dpdm.html}} (\bibinfo {year} {2025})\BibitemShut {NoStop}%
\bibitem [{\citenamefont {{Cohen}}\ \emph {et~al.}(1950)\citenamefont {{Cohen}}, \citenamefont {{Spitzer}},\ and\ \citenamefont {{Routly}}}]{1950PhRv...80..230C}%
  \BibitemOpen
  \bibfield  {author} {\bibinfo {author} {\bibfnamefont {R.~S.}\ \bibnamefont {{Cohen}}}, \bibinfo {author} {\bibfnamefont {L.}~\bibnamefont {{Spitzer}}},\ and\ \bibinfo {author} {\bibfnamefont {P.~M.}\ \bibnamefont {{Routly}}},\ }\href {https://doi.org/10.1103/PhysRev.80.230} {\bibfield  {journal} {\bibinfo  {journal} {Physical Review}\ }\textbf {\bibinfo {volume} {80}},\ \bibinfo {pages} {230} (\bibinfo {year} {1950})}\BibitemShut {NoStop}%
\bibitem [{\citenamefont {{Spitzer}}(1962)}]{1962pfig.book.....S}%
  \BibitemOpen
  \bibfield  {author} {\bibinfo {author} {\bibfnamefont {L.}~\bibnamefont {{Spitzer}}},\ }\href@noop {} {\emph {\bibinfo {title} {{Physics of Fully Ionized Gases}}}}\ (\bibinfo {year} {1962})\BibitemShut {NoStop}%
\bibitem [{\citenamefont {{Tigik}}\ \emph {et~al.}(2016)\citenamefont {{Tigik}}, \citenamefont {{Ziebell}},\ and\ \citenamefont {{Yoon}}}]{2016PhPl...23f4504T}%
  \BibitemOpen
  \bibfield  {author} {\bibinfo {author} {\bibfnamefont {S.~F.}\ \bibnamefont {{Tigik}}}, \bibinfo {author} {\bibfnamefont {L.~F.}\ \bibnamefont {{Ziebell}}},\ and\ \bibinfo {author} {\bibfnamefont {P.~H.}\ \bibnamefont {{Yoon}}},\ }\href {https://doi.org/10.1063/1.4953802} {\bibfield  {journal} {\bibinfo  {journal} {Physics of Plasmas}\ }\textbf {\bibinfo {volume} {23}},\ \bibinfo {eid} {064504} (\bibinfo {year} {2016})},\ \Eprint {https://arxiv.org/abs/1710.03874} {arXiv:1710.03874 [physics.plasm-ph]} \BibitemShut {NoStop}%
\bibitem [{\citenamefont {Dubovsky}\ and\ \citenamefont {Hern{\'a}ndez-Chifflet}(2015)}]{Dubovsky:2015cca}%
  \BibitemOpen
  \bibfield  {author} {\bibinfo {author} {\bibfnamefont {S.}~\bibnamefont {Dubovsky}}\ and\ \bibinfo {author} {\bibfnamefont {G.}~\bibnamefont {Hern{\'a}ndez-Chifflet}},\ }\href {https://doi.org/10.1088/1475-7516/2015/12/054} {\bibfield  {journal} {\bibinfo  {journal} {JCAP}\ }\textbf {\bibinfo {volume} {12}},\ \bibinfo {pages} {054}},\ \Eprint {https://arxiv.org/abs/1509.00039} {arXiv:1509.00039 [hep-ph]} \BibitemShut {NoStop}%
\bibitem [{\citenamefont {Liu}\ \emph {et~al.}(2021)\citenamefont {Liu}, \citenamefont {Qin}, \citenamefont {Ridgway},\ and\ \citenamefont {Slatyer}}]{Liu:2020wqz}%
  \BibitemOpen
  \bibfield  {author} {\bibinfo {author} {\bibfnamefont {H.}~\bibnamefont {Liu}}, \bibinfo {author} {\bibfnamefont {W.}~\bibnamefont {Qin}}, \bibinfo {author} {\bibfnamefont {G.~W.}\ \bibnamefont {Ridgway}},\ and\ \bibinfo {author} {\bibfnamefont {T.~R.}\ \bibnamefont {Slatyer}},\ }\href {https://doi.org/10.1103/PhysRevD.104.043514} {\bibfield  {journal} {\bibinfo  {journal} {Phys. Rev. D}\ }\textbf {\bibinfo {volume} {104}},\ \bibinfo {pages} {043514} (\bibinfo {year} {2021})},\ \Eprint {https://arxiv.org/abs/2008.01084} {arXiv:2008.01084 [astro-ph.CO]} \BibitemShut {NoStop}%
\bibitem [{\citenamefont {Abac}\ \emph {et~al.}(2025)\citenamefont {Abac} \emph {et~al.}}]{LIGOScientific:2025csr}%
  \BibitemOpen
  \bibfield  {author} {\bibinfo {author} {\bibfnamefont {A.~G.}\ \bibnamefont {Abac}} \emph {et~al.} (\bibinfo {collaboration} {LIGO Scientific, VIRGO, KAGRA}),\ }\href@noop {} {\  (\bibinfo {year} {2025})},\ \Eprint {https://arxiv.org/abs/2509.07352} {arXiv:2509.07352 [gr-qc]} \BibitemShut {NoStop}%
\bibitem [{\citenamefont {Aswathi}\ \emph {et~al.}(2025)\citenamefont {Aswathi}, \citenamefont {East}, \citenamefont {Siemonsen}, \citenamefont {Sun},\ and\ \citenamefont {Jones}}]{Aswathi:2025nxa}%
  \BibitemOpen
  \bibfield  {author} {\bibinfo {author} {\bibfnamefont {P.~S.}\ \bibnamefont {Aswathi}}, \bibinfo {author} {\bibfnamefont {W.~E.}\ \bibnamefont {East}}, \bibinfo {author} {\bibfnamefont {N.}~\bibnamefont {Siemonsen}}, \bibinfo {author} {\bibfnamefont {L.}~\bibnamefont {Sun}},\ and\ \bibinfo {author} {\bibfnamefont {D.}~\bibnamefont {Jones}},\ }\href@noop {} {\  (\bibinfo {year} {2025})},\ \Eprint {https://arxiv.org/abs/2507.20979} {arXiv:2507.20979 [gr-qc]} \BibitemShut {NoStop}%
\bibitem [{\citenamefont {Hook}\ \emph {et~al.}(2018)\citenamefont {Hook}, \citenamefont {Kahn}, \citenamefont {Safdi},\ and\ \citenamefont {Sun}}]{Hook:2018iia}%
  \BibitemOpen
  \bibfield  {author} {\bibinfo {author} {\bibfnamefont {A.}~\bibnamefont {Hook}}, \bibinfo {author} {\bibfnamefont {Y.}~\bibnamefont {Kahn}}, \bibinfo {author} {\bibfnamefont {B.~R.}\ \bibnamefont {Safdi}},\ and\ \bibinfo {author} {\bibfnamefont {Z.}~\bibnamefont {Sun}},\ }\href {https://doi.org/10.1103/PhysRevLett.121.241102} {\bibfield  {journal} {\bibinfo  {journal} {Phys. Rev. Lett.}\ }\textbf {\bibinfo {volume} {121}},\ \bibinfo {pages} {241102} (\bibinfo {year} {2018})},\ \Eprint {https://arxiv.org/abs/1804.03145} {arXiv:1804.03145 [hep-ph]} \BibitemShut {NoStop}%
\bibitem [{\citenamefont {An}\ \emph {et~al.}(2025)\citenamefont {An}, \citenamefont {Ge}, \citenamefont {Liu},\ and\ \citenamefont {Liu}}]{An:2024wmc}%
  \BibitemOpen
  \bibfield  {author} {\bibinfo {author} {\bibfnamefont {H.}~\bibnamefont {An}}, \bibinfo {author} {\bibfnamefont {S.}~\bibnamefont {Ge}}, \bibinfo {author} {\bibfnamefont {J.}~\bibnamefont {Liu}},\ and\ \bibinfo {author} {\bibfnamefont {M.}~\bibnamefont {Liu}},\ }\href {https://doi.org/10.1103/PhysRevLett.134.171001} {\bibfield  {journal} {\bibinfo  {journal} {Phys. Rev. Lett.}\ }\textbf {\bibinfo {volume} {134}},\ \bibinfo {pages} {171001} (\bibinfo {year} {2025})},\ \Eprint {https://arxiv.org/abs/2405.12285} {arXiv:2405.12285 [hep-ph]} \BibitemShut {NoStop}%
\bibitem [{\citenamefont {Fox}\ \emph {et~al.}(2025)\citenamefont {Fox}, \citenamefont {Weiner},\ and\ \citenamefont {Xiao}}]{Fox:2025tqa}%
  \BibitemOpen
  \bibfield  {author} {\bibinfo {author} {\bibfnamefont {P.~J.}\ \bibnamefont {Fox}}, \bibinfo {author} {\bibfnamefont {N.}~\bibnamefont {Weiner}},\ and\ \bibinfo {author} {\bibfnamefont {H.}~\bibnamefont {Xiao}},\ }\href@noop {} {\  (\bibinfo {year} {2025})},\ \Eprint {https://arxiv.org/abs/2508.08371} {arXiv:2508.08371 [hep-ph]} \BibitemShut {NoStop}%
\bibitem [{\citenamefont {{Vay}}(2008)}]{Vay-2008}%
  \BibitemOpen
  \bibfield  {author} {\bibinfo {author} {\bibfnamefont {J.~L.}\ \bibnamefont {{Vay}}},\ }\href {https://doi.org/10.1063/1.2837054} {\bibfield  {journal} {\bibinfo  {journal} {Physics of Plasmas}\ }\textbf {\bibinfo {volume} {15}},\ \bibinfo {eid} {056701} (\bibinfo {year} {2008})}\BibitemShut {NoStop}%
\bibitem [{\citenamefont {{Shalaby}}\ \emph {et~al.}(2017{\natexlab{b}})\citenamefont {{Shalaby}}, \citenamefont {{Broderick}}, \citenamefont {{Chang}}, \citenamefont {{Pfrommer}}, \citenamefont {{Lamberts}},\ and\ \citenamefont {{Puchwein}}}]{resolution-paper}%
  \BibitemOpen
  \bibfield  {author} {\bibinfo {author} {\bibfnamefont {M.}~\bibnamefont {{Shalaby}}}, \bibinfo {author} {\bibfnamefont {A.~E.}\ \bibnamefont {{Broderick}}}, \bibinfo {author} {\bibfnamefont {P.}~\bibnamefont {{Chang}}}, \bibinfo {author} {\bibfnamefont {C.}~\bibnamefont {{Pfrommer}}}, \bibinfo {author} {\bibfnamefont {A.}~\bibnamefont {{Lamberts}}},\ and\ \bibinfo {author} {\bibfnamefont {E.}~\bibnamefont {{Puchwein}}},\ }\href {https://doi.org/10.3847/1538-4357/aa8b17} {\bibfield  {journal} {\bibinfo  {journal} {\apj}\ }\textbf {\bibinfo {volume} {848}},\ \bibinfo {eid} {81} (\bibinfo {year} {2017}{\natexlab{b}})},\ \Eprint {https://arxiv.org/abs/1704.00014} {arXiv:1704.00014 [astro-ph.HE]} \BibitemShut {NoStop}%
\bibitem [{\citenamefont {{Shalaby}}\ \emph {et~al.}(2018)\citenamefont {{Shalaby}}, \citenamefont {{Broderick}}, \citenamefont {{Chang}}, \citenamefont {{Pfrommer}}, \citenamefont {{Lamberts}},\ and\ \citenamefont {{Puchwein}}}]{sim_inho_18}%
  \BibitemOpen
  \bibfield  {author} {\bibinfo {author} {\bibfnamefont {M.}~\bibnamefont {{Shalaby}}}, \bibinfo {author} {\bibfnamefont {A.~E.}\ \bibnamefont {{Broderick}}}, \bibinfo {author} {\bibfnamefont {P.}~\bibnamefont {{Chang}}}, \bibinfo {author} {\bibfnamefont {C.}~\bibnamefont {{Pfrommer}}}, \bibinfo {author} {\bibfnamefont {A.}~\bibnamefont {{Lamberts}}},\ and\ \bibinfo {author} {\bibfnamefont {E.}~\bibnamefont {{Puchwein}}},\ }\href {https://doi.org/10.3847/1538-4357/aabe92} {\bibfield  {journal} {\bibinfo  {journal} {\apj}\ }\textbf {\bibinfo {volume} {859}},\ \bibinfo {eid} {45} (\bibinfo {year} {2018})},\ \Eprint {https://arxiv.org/abs/1804.05071} {arXiv:1804.05071 [astro-ph.HE]} \BibitemShut {NoStop}%
\bibitem [{\citenamefont {{Shalaby}}\ \emph {et~al.}(2020)\citenamefont {{Shalaby}}, \citenamefont {{Broderick}}, \citenamefont {{Chang}}, \citenamefont {{Pfrommer}}, \citenamefont {{Puchwein}},\ and\ \citenamefont {{Lamberts}}}]{th_inho_20}%
  \BibitemOpen
  \bibfield  {author} {\bibinfo {author} {\bibfnamefont {M.}~\bibnamefont {{Shalaby}}}, \bibinfo {author} {\bibfnamefont {A.~E.}\ \bibnamefont {{Broderick}}}, \bibinfo {author} {\bibfnamefont {P.}~\bibnamefont {{Chang}}}, \bibinfo {author} {\bibfnamefont {C.}~\bibnamefont {{Pfrommer}}}, \bibinfo {author} {\bibfnamefont {E.}~\bibnamefont {{Puchwein}}},\ and\ \bibinfo {author} {\bibfnamefont {A.}~\bibnamefont {{Lamberts}}},\ }\href {https://doi.org/10.1017/S0022377820000215} {\bibfield  {journal} {\bibinfo  {journal} {Journal of Plasma Physics}\ }\textbf {\bibinfo {volume} {86}},\ \bibinfo {eid} {535860201} (\bibinfo {year} {2020})},\ \Eprint {https://arxiv.org/abs/2003.02849} {arXiv:2003.02849 [astro-ph.HE]} \BibitemShut {NoStop}%
\bibitem [{\citenamefont {{Lemmerz}}\ \emph {et~al.}(2024)\citenamefont {{Lemmerz}}, \citenamefont {{Shalaby}}, \citenamefont {{Thomas}},\ and\ \citenamefont {{Pfrommer}}}]{Lemmerz+2023}%
  \BibitemOpen
  \bibfield  {author} {\bibinfo {author} {\bibfnamefont {R.}~\bibnamefont {{Lemmerz}}}, \bibinfo {author} {\bibfnamefont {M.}~\bibnamefont {{Shalaby}}}, \bibinfo {author} {\bibfnamefont {T.}~\bibnamefont {{Thomas}}},\ and\ \bibinfo {author} {\bibfnamefont {C.}~\bibnamefont {{Pfrommer}}},\ }\href {https://doi.org/10.1017/S0022377823001113} {\bibfield  {journal} {\bibinfo  {journal} {Journal of Plasma Physics}\ }\textbf {\bibinfo {volume} {90}},\ \bibinfo {eid} {905900104} (\bibinfo {year} {2024})},\ \Eprint {https://arxiv.org/abs/2301.04679} {arXiv:2301.04679 [astro-ph.HE]} \BibitemShut {NoStop}%
\bibitem [{\citenamefont {{Lemmerz}}\ \emph {et~al.}(2025)\citenamefont {{Lemmerz}}, \citenamefont {{Shalaby}}, \citenamefont {{Pfrommer}},\ and\ \citenamefont {{Thomas}}}]{Lemmerz2025}%
  \BibitemOpen
  \bibfield  {author} {\bibinfo {author} {\bibfnamefont {R.}~\bibnamefont {{Lemmerz}}}, \bibinfo {author} {\bibfnamefont {M.}~\bibnamefont {{Shalaby}}}, \bibinfo {author} {\bibfnamefont {C.}~\bibnamefont {{Pfrommer}}},\ and\ \bibinfo {author} {\bibfnamefont {T.}~\bibnamefont {{Thomas}}},\ }\href {https://doi.org/10.3847/1538-4357/ad8eb3} {\bibfield  {journal} {\bibinfo  {journal} {\apj}\ }\textbf {\bibinfo {volume} {979}},\ \bibinfo {eid} {34} (\bibinfo {year} {2025})},\ \Eprint {https://arxiv.org/abs/2406.04400} {arXiv:2406.04400 [astro-ph.HE]} \BibitemShut {NoStop}%
\bibitem [{\citenamefont {{Shalaby}}\ \emph {et~al.}(2022)\citenamefont {{Shalaby}}, \citenamefont {{Lemmerz}}, \citenamefont {{Thomas}},\ and\ \citenamefont {{Pfrommer}}}]{Shalaby+2022ApJ}%
  \BibitemOpen
  \bibfield  {author} {\bibinfo {author} {\bibfnamefont {M.}~\bibnamefont {{Shalaby}}}, \bibinfo {author} {\bibfnamefont {R.}~\bibnamefont {{Lemmerz}}}, \bibinfo {author} {\bibfnamefont {T.}~\bibnamefont {{Thomas}}},\ and\ \bibinfo {author} {\bibfnamefont {C.}~\bibnamefont {{Pfrommer}}},\ }\href {https://doi.org/10.3847/1538-4357/ac6ce7} {\bibfield  {journal} {\bibinfo  {journal} {\apj}\ }\textbf {\bibinfo {volume} {932}},\ \bibinfo {eid} {86} (\bibinfo {year} {2022})},\ \Eprint {https://arxiv.org/abs/2202.05288} {arXiv:2202.05288 [astro-ph.HE]} \BibitemShut {NoStop}%
\bibitem [{\citenamefont {{Shalaby}}(2024)}]{Shalaby2024ApJL}%
  \BibitemOpen
  \bibfield  {author} {\bibinfo {author} {\bibfnamefont {M.}~\bibnamefont {{Shalaby}}},\ }\href {https://doi.org/10.3847/2041-8213/ad99d8} {\bibfield  {journal} {\bibinfo  {journal} {\apjl}\ }\textbf {\bibinfo {volume} {977}},\ \bibinfo {eid} {L43} (\bibinfo {year} {2024})},\ \Eprint {https://arxiv.org/abs/2412.03530} {arXiv:2412.03530 [astro-ph.HE]} \BibitemShut {NoStop}%
\bibitem [{\citenamefont {{Shalaby}}\ \emph {et~al.}(2025)\citenamefont {{Shalaby}}, \citenamefont {{Bret}},\ and\ \citenamefont {{Fraschetti}}}]{Shalaby2025ApJ}%
  \BibitemOpen
  \bibfield  {author} {\bibinfo {author} {\bibfnamefont {M.}~\bibnamefont {{Shalaby}}}, \bibinfo {author} {\bibfnamefont {A.}~\bibnamefont {{Bret}}},\ and\ \bibinfo {author} {\bibfnamefont {F.}~\bibnamefont {{Fraschetti}}},\ }\href {https://doi.org/10.3847/1538-4357/adf84f} {\bibfield  {journal} {\bibinfo  {journal} {\apj}\ }\textbf {\bibinfo {volume} {991}},\ \bibinfo {eid} {26} (\bibinfo {year} {2025})},\ \Eprint {https://arxiv.org/abs/2503.10758} {arXiv:2503.10758 [astro-ph.SR]} \BibitemShut {NoStop}%
\bibitem [{\citenamefont {{Birdsall}}\ and\ \citenamefont {{Maron}}(1980)}]{birdsall+1980}%
  \BibitemOpen
  \bibfield  {author} {\bibinfo {author} {\bibfnamefont {C.~K.}\ \bibnamefont {{Birdsall}}}\ and\ \bibinfo {author} {\bibfnamefont {N.}~\bibnamefont {{Maron}}},\ }\href {https://doi.org/10.1016/0021-9991(80)90171-0} {\bibfield  {journal} {\bibinfo  {journal} {Journal of Computational Physics}\ }\textbf {\bibinfo {volume} {36}},\ \bibinfo {pages} {1} (\bibinfo {year} {1980})}\BibitemShut {NoStop}%
\bibitem [{\citenamefont {Sailer}\ \emph {et~al.}(2025)\citenamefont {Sailer}, \citenamefont {Farren}, \citenamefont {Ferraro},\ and\ \citenamefont {White}}]{Sailer:2025lxj}%
  \BibitemOpen
  \bibfield  {author} {\bibinfo {author} {\bibfnamefont {N.}~\bibnamefont {Sailer}}, \bibinfo {author} {\bibfnamefont {G.~S.}\ \bibnamefont {Farren}}, \bibinfo {author} {\bibfnamefont {S.}~\bibnamefont {Ferraro}},\ and\ \bibinfo {author} {\bibfnamefont {M.}~\bibnamefont {White}},\ }\href@noop {} {\  (\bibinfo {year} {2025})},\ \Eprint {https://arxiv.org/abs/2504.16932} {arXiv:2504.16932 [astro-ph.CO]} \BibitemShut {NoStop}%
\bibitem [{\citenamefont {Cheng}\ \emph {et~al.}(2025)\citenamefont {Cheng}, \citenamefont {Yin}, \citenamefont {Di~Valentino}, \citenamefont {Marsh},\ and\ \citenamefont {Visinelli}}]{Cheng:2025cmb}%
  \BibitemOpen
  \bibfield  {author} {\bibinfo {author} {\bibfnamefont {H.}~\bibnamefont {Cheng}}, \bibinfo {author} {\bibfnamefont {Z.}~\bibnamefont {Yin}}, \bibinfo {author} {\bibfnamefont {E.}~\bibnamefont {Di~Valentino}}, \bibinfo {author} {\bibfnamefont {D.~J.~E.}\ \bibnamefont {Marsh}},\ and\ \bibinfo {author} {\bibfnamefont {L.}~\bibnamefont {Visinelli}},\ }\href@noop {} {\  (\bibinfo {year} {2025})},\ \Eprint {https://arxiv.org/abs/2506.19096} {arXiv:2506.19096 [astro-ph.CO]} \BibitemShut {NoStop}%
\bibitem [{\citenamefont {Trost}\ \emph {et~al.}(2025)\citenamefont {Trost}, \citenamefont {Bolton}, \citenamefont {Caputo}, \citenamefont {Liu}, \citenamefont {Cristiani},\ and\ \citenamefont {Viel}}]{Trost:2024ciu}%
  \BibitemOpen
  \bibfield  {author} {\bibinfo {author} {\bibfnamefont {A.}~\bibnamefont {Trost}}, \bibinfo {author} {\bibfnamefont {J.~S.}\ \bibnamefont {Bolton}}, \bibinfo {author} {\bibfnamefont {A.}~\bibnamefont {Caputo}}, \bibinfo {author} {\bibfnamefont {H.}~\bibnamefont {Liu}}, \bibinfo {author} {\bibfnamefont {S.}~\bibnamefont {Cristiani}},\ and\ \bibinfo {author} {\bibfnamefont {M.}~\bibnamefont {Viel}},\ }\href {https://doi.org/10.1103/PhysRevD.111.083034} {\bibfield  {journal} {\bibinfo  {journal} {Phys. Rev. D}\ }\textbf {\bibinfo {volume} {111}},\ \bibinfo {pages} {083034} (\bibinfo {year} {2025})},\ \Eprint {https://arxiv.org/abs/2410.02858} {arXiv:2410.02858 [astro-ph.CO]} \BibitemShut {NoStop}%
\bibitem [{\citenamefont {Slatyer}(2016)}]{Slatyer:2015kla}%
  \BibitemOpen
  \bibfield  {author} {\bibinfo {author} {\bibfnamefont {T.~R.}\ \bibnamefont {Slatyer}},\ }\href {https://doi.org/10.1103/PhysRevD.93.023521} {\bibfield  {journal} {\bibinfo  {journal} {Phys. Rev. D}\ }\textbf {\bibinfo {volume} {93}},\ \bibinfo {pages} {023521} (\bibinfo {year} {2016})},\ \Eprint {https://arxiv.org/abs/1506.03812} {arXiv:1506.03812 [astro-ph.CO]} \BibitemShut {NoStop}%
\bibitem [{\citenamefont {Chluba}\ and\ \citenamefont {Sunyaev}(2012)}]{Chluba:2011hw}%
  \BibitemOpen
  \bibfield  {author} {\bibinfo {author} {\bibfnamefont {J.}~\bibnamefont {Chluba}}\ and\ \bibinfo {author} {\bibfnamefont {R.~A.}\ \bibnamefont {Sunyaev}},\ }\href {https://doi.org/10.1111/j.1365-2966.2011.19786.x} {\bibfield  {journal} {\bibinfo  {journal} {Mon. Not. Roy. Astron. Soc.}\ }\textbf {\bibinfo {volume} {419}},\ \bibinfo {pages} {1294} (\bibinfo {year} {2012})},\ \Eprint {https://arxiv.org/abs/1109.6552} {arXiv:1109.6552 [astro-ph.CO]} \BibitemShut {NoStop}%
\bibitem [{\citenamefont {Hill}\ \emph {et~al.}(2015)\citenamefont {Hill}, \citenamefont {Battaglia}, \citenamefont {Chluba}, \citenamefont {Ferraro}, \citenamefont {Schaan},\ and\ \citenamefont {Spergel}}]{Hill:2015tqa}%
  \BibitemOpen
  \bibfield  {author} {\bibinfo {author} {\bibfnamefont {J.~C.}\ \bibnamefont {Hill}}, \bibinfo {author} {\bibfnamefont {N.}~\bibnamefont {Battaglia}}, \bibinfo {author} {\bibfnamefont {J.}~\bibnamefont {Chluba}}, \bibinfo {author} {\bibfnamefont {S.}~\bibnamefont {Ferraro}}, \bibinfo {author} {\bibfnamefont {E.}~\bibnamefont {Schaan}},\ and\ \bibinfo {author} {\bibfnamefont {D.~N.}\ \bibnamefont {Spergel}},\ }\href {https://doi.org/10.1103/PhysRevLett.115.261301} {\bibfield  {journal} {\bibinfo  {journal} {Phys. Rev. Lett.}\ }\textbf {\bibinfo {volume} {115}},\ \bibinfo {pages} {261301} (\bibinfo {year} {2015})},\ \Eprint {https://arxiv.org/abs/1507.01583} {arXiv:1507.01583 [astro-ph.CO]} \BibitemShut {NoStop}%
\bibitem [{\citenamefont {An}\ \emph {et~al.}(2013)\citenamefont {An}, \citenamefont {Pospelov},\ and\ \citenamefont {Pradler}}]{An:2013yfc}%
  \BibitemOpen
  \bibfield  {author} {\bibinfo {author} {\bibfnamefont {H.}~\bibnamefont {An}}, \bibinfo {author} {\bibfnamefont {M.}~\bibnamefont {Pospelov}},\ and\ \bibinfo {author} {\bibfnamefont {J.}~\bibnamefont {Pradler}},\ }\href {https://doi.org/10.1016/j.physletb.2013.07.008} {\bibfield  {journal} {\bibinfo  {journal} {Phys. Lett.}\ }\textbf {\bibinfo {volume} {B725}},\ \bibinfo {pages} {190} (\bibinfo {year} {2013})},\ \Eprint {https://arxiv.org/abs/1302.3884} {arXiv:1302.3884 [hep-ph]} \BibitemShut {NoStop}%
\bibitem [{\citenamefont {Wadekar}\ and\ \citenamefont {Farrar}(2021)}]{Wadekar:2019mpc}%
  \BibitemOpen
  \bibfield  {author} {\bibinfo {author} {\bibfnamefont {D.}~\bibnamefont {Wadekar}}\ and\ \bibinfo {author} {\bibfnamefont {G.~R.}\ \bibnamefont {Farrar}},\ }\href {https://doi.org/10.1103/PhysRevD.103.123028} {\bibfield  {journal} {\bibinfo  {journal} {Phys. Rev. D}\ }\textbf {\bibinfo {volume} {103}},\ \bibinfo {pages} {123028} (\bibinfo {year} {2021})},\ \Eprint {https://arxiv.org/abs/1903.12190} {arXiv:1903.12190 [hep-ph]} \BibitemShut {NoStop}%
\bibitem [{\citenamefont {Coulton}\ \emph {et~al.}(2020)\citenamefont {Coulton}, \citenamefont {Ota},\ and\ \citenamefont {van Engelen}}]{Coulton:2019ign}%
  \BibitemOpen
  \bibfield  {author} {\bibinfo {author} {\bibfnamefont {W.~R.}\ \bibnamefont {Coulton}}, \bibinfo {author} {\bibfnamefont {A.}~\bibnamefont {Ota}},\ and\ \bibinfo {author} {\bibfnamefont {A.}~\bibnamefont {van Engelen}},\ }\href {https://doi.org/10.1103/PhysRevLett.125.111301} {\bibfield  {journal} {\bibinfo  {journal} {Phys. Rev. Lett.}\ }\textbf {\bibinfo {volume} {125}},\ \bibinfo {pages} {111301} (\bibinfo {year} {2020})},\ \Eprint {https://arxiv.org/abs/1910.10152} {arXiv:1910.10152 [astro-ph.CO]} \BibitemShut {NoStop}%
\bibitem [{\citenamefont {Sun}\ \emph {et~al.}(2025)\citenamefont {Sun}, \citenamefont {Foster},\ and\ \citenamefont {Mu{\~n}oz}}]{Sun:2025ksr}%
  \BibitemOpen
  \bibfield  {author} {\bibinfo {author} {\bibfnamefont {Y.}~\bibnamefont {Sun}}, \bibinfo {author} {\bibfnamefont {J.~W.}\ \bibnamefont {Foster}},\ and\ \bibinfo {author} {\bibfnamefont {J.~B.}\ \bibnamefont {Mu{\~n}oz}},\ }\href@noop {} {\  (\bibinfo {year} {2025})},\ \Eprint {https://arxiv.org/abs/2509.22772} {arXiv:2509.22772 [hep-ph]} \BibitemShut {NoStop}%
\end{thebibliography}%
\end{document}